\documentclass[manuscript=article]{achemso}
\usepackage{graphicx}
\usepackage{xcolor,soul}
\usepackage[colorlinks=true,linkcolor=blue,urlcolor=blue,citecolor=blue]{hyperref}
\usepackage[font={sf,small},labelfont=bf]{caption}
\usepackage{amsmath}
\usepackage{booktabs}
\setcitestyle{square,numbers,sort&compress}

\newcommand{\thetitle}{A scalable method for cavity--enhanced solid--state quantum sensors}

\title{\thetitle} 

\author{Daniel J. Tibben}
\email{dtibben1@gmail.com}
\altaffiliation{These authors contributed equally to this work.}
\author{Roy Styles}
\altaffiliation{These authors contributed equally to this work.}
\author{David A. Broadway}
\author{Jean-Philippe Tetienne}
\author{Daniel E. G\'omez}
\email{daniel.gomez@rmit.edu.au}
\author{Philipp Reineck}
\email{philipp.reineck@rmit.edu.au}
\affiliation{School of Science, RMIT University, Melbourne, Australia}

\date{\today}

\begin{document}

\maketitle

\begin{abstract}
Photoluminescent color centers in diamond and hexagonal boron nitride (hBN) are powerful nanoscale solid-state quantum sensors that are explored in a plethora of quantum technologies. Methods for integrating them into macroscopic structures that improve their sensitivity and enable their large-scale deployment are highly sought after. Here, we demonstrate cavity-enhanced photoluminescence (PL) of fluorescent nanodiamonds (FNDs) and hBN nanoparticles (NPs) embedded in polymer-based thin-film optical cavities on the centimeter scale. The cavity resonances efficiently modulate the spectral PL peak position of nitrogen-vacancy (NV) centers in FNDs across the NV PL spectrum and lead to an up to 2.9-fold Purcell-enhancement of the NV PL decay rate. The brightness of hBN NPs increases by up to a factor of three and the PL decay rate is enhanced by up to 13-fold inside the cavities. Finally, we find a 4.8 times improved magnetic field sensitivity of 20 nm FNDs in thin-film cavities due to cavity-enhanced optically detected magnetic resonance contrast and PL brightness. Our study demonstrates a low-cost and scalable method for the fabrication of quantum sensor-doped thin-film cavities, which is an important step toward the development of advanced quantum sensing technologies.
\end{abstract}


\section{Introduction}
Photoluminescent defects, in diamond, hexagonal boron nitride and silicon carbide with optically addressable spins are vital for a broad range of emerging room-temperature quantum technologies. 
The nitrogen vacancy (NV) center in diamond is the most technologically advanced defect and has been used for many imaging and sensing applications~\cite{Rovny_NRP2024,Aslam_NRP2023}, most of which are based on bulk diamond samples. 
However, diamond is hard and brittle, relatively expensive, and today's quantum diamond chips are only available up to 4~mm in size. 
This limits their integration with other materials and their use in large-area or high-throughput (\textit{e.g.} point-of-care biomedical devices) applications. 
Hence, diamond nanoparticles (NPs) on solid substrates~\cite{Heffernan_SR2017,Kianinia_N2016,Shulevitz_AN2022,Foy_AAMI2020,Chea_2025}
or integrated into other materials~\cite{Andrich_NL2018,Styles_AOM2024,Guarino_AANM2020,Khalid_AAMI2020,Price_S2019}
are explored for these applications. However, while fluorescent nanodiamonds (FNDs) containing NV centers offer many potential advantages, their photoluminescence (PL) brightness is generally lower and their spin coherence times shorter than that of NVs in bulk diamond, limiting their sensitivity. 

More recently, spin-active defects in hBN have emerged as a promising 2D material platform for nanoscale sensing~\cite{Aharonovich_NL2022}. 
Unlike diamond, many as-synthesized hBN materials from NPs to bulk crystals, contain a range of quantum emitters with PL from the violet to the near-infrared spectral region in the same material~\cite{Singh_AM2025}. 
However, the PL properties of individual particles can vary greatly~\cite{Xu_N2018} and one defect that can be controllably created in hBN~\cite{Choi_AAMI2016}, the boron vacancy $V_B^-$~\cite{Tran_NN2016,Gottscholl_NM2020}, is very dim and has not been observed as a single photon emitter.  

Hence, methods for controlling and enhancing FND and hBN PL properties are highly sought after. 
One approach is to use optical cavities for Purcell-enhanced PL, which has been explored for emitters in diamond~\cite{Janitz_OO2020,Katsumi_CE2025}
and hBN~\cite{Jang_AM2023,Wang_PSSB2025}.
For diamond, this has been explored for individual FNDs in photonic crystal cavities~\cite{Wolters_APL2010,Tomljenovic-Hanic_OEO2011,Fehler_AN2019,Schrinner_NL2020},
in metamaterial photonic cavities~\cite{Bar-David_PRA2023}, 
in open cavities~\cite{Kaupp_PRA2016,Jeske_NC2017}, 
and photonic crystal cavities in bulk diamond~\cite{Liang_OFCC2Po2012,Riedel_PRX2017,Schroder_C22PF2014,Berghaus_PRA2025}.
For hBN, this enhance ment approach has been explored in planar Fabry-P\'erot--based cavities~\cite{Vogl_AP2019,Zeng_N2023,Scheuer_AOA2023}, 
open cavities~\cite{Haussler_AOM2021}, 
plasmonic cavities~\cite{Mendelson_NM2021,Genc_2025}, 
micronanoresonators~\cite{Proscia_N2020,Parto_NL2022,Sakib_NL2024}, 
and photonic crystal cavities~\cite{Froch_S2022,Nonahal_NL2023,Qian_NL2022,Froch_AN2020,Kim_NC2018}. 
However, to date, no straightforward and scalable method for the fabrication of thin-film cavities containing nanoscale quantum sensors has been reported. 
Wafer-scale quantum sensor-doped thin-film cavities can enable, for example, magnetic imaging of currents in microelectronics~\cite{Turner_PRA2020} as illustrated in Fig.~\ref{fig:scheme}a.

Here, we demonstrate cavity-enhanced PL of FND and hBN NP quantum sensors embedded in polymer-based thin-film cavities. 
We report a straightforward and scalable method for the fabrication of quantum sensor-doped thin-film microcavities on the centimeter scale. 
The PL properties of FNDs and hBN NPs are investigated as a function of the spectral position of the cavity resonance across the visible and near-infrared spectral range. 
The cavity resonances efficiently modulate the spectral PL peak position of NV centers in FNDs across the NV PL spectrum and lead to an up to 2.9-fold enhancement of the NV PL decay rate compared to FNDs outside the cavity. 
The hBN NP PL brightness increases by up to a factor of three and the PL decay rate is enhanced up to 13-fold inside the cavity thin-films. 
By comparing experimental and theoretical Purcell enhancement factors, we conclude that Purcell enhancement causes the observed modulation of NV PL and that Purcell enhancement alone cannot explain the strong PL brightness and decay enhancement of hBN NP PL. 
Finally, we demonstrate an up to 4.8 times improved magnetic field sensitivity of 20~nm and 100~nm FNDs in thin-film cavities due to a cavity-enhanced optically detected magnetic resonance (ODMR) contrast.

\section{Results and Discussion}

\begin{figure}[t!]
    \includegraphics[width=\linewidth]{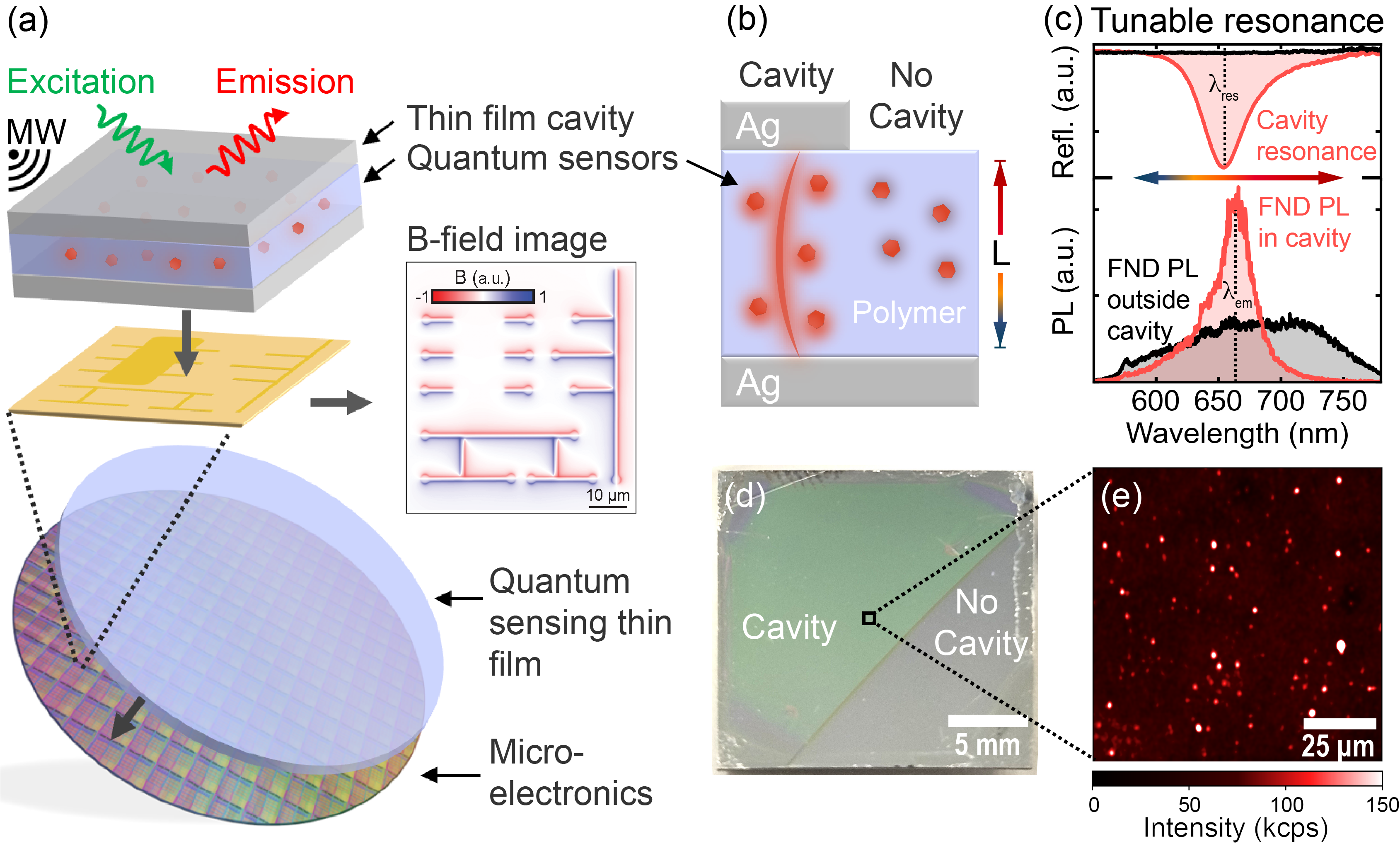}
    \caption{\textbf{Application and design of quantum sensor-doped microcavity thin-films---} 
    (a) Schematic illustration of the application of our thin-films to the magnetic imaging of currents in a microcircuit \textit{via} cavity--enhanced quantum sensors.
    (b) Schematic of the cavity and control ('no cavity') devices with embedded quantum sensor particles. 
    The cavity resonance energy is controlled by the polymer layer thickness $L$, allowing tuning of the spectral position of the cavity resonance.
    (c) Typical reflectivity spectrum of the cavity resonance (top) and PL spectra of FNDs inside (red trace) and outside the cavity (black trace). 
    (d) A photograph of a typical thin-film microcavity-coated substrate, with clearly distinguishable cavity (green) and no cavity (clear) regions. 
    (e)  Confocal PL image of the microcavity in panel (c), showing PL from FNDs uniformly dispersed throughout the thin-film. 
    }
    \label{fig:scheme}
\end{figure}

Low-cost device-integrated quantum sensors, such as the thin-film of quantum sensing depicted in Fig.~\ref{fig:scheme}a, are critical to their widespread adoption in commercial applications.  
To this end, we designed a series of polymer-based Fabry-P\'erot microcavity devices with embedded FNDs and hBN NPs and explored the cavity--modified optical properties of the quantum sensors by comparing their optical properties inside and outside the cavity on the same device, as illustrated in Figure Fig~\ref{fig:scheme}b. 
The devices comprised of a silver mirror on either a Si wafer or quartz substrate as the base layer, followed by a polymer thin-film doped with either FNDs or hBN NPs, and capped with a semi-transparent top mirror. 
This design ensures the formation of confined optical cavity modes that interact with the quantum sensors embedded in the polymer layer. 
The resonance wavelength of the modes can be controlled \textit{via} the polymer thickness $L$. 
As an example, Fig~\ref{fig:scheme}c, top,  shows the reflectance spectrum  of a cavity with a polymer layer thickness of 159~nm and a typical PL spectrum of an FND inside this cavity (bottom). 
The cavity shows a pronounced resonance peak at $\lambda_{res} = 650$~nm. 
This resonance leads to a spectrally narrower cavity-enhanced FND PL peaking at $\lambda_{em} = 664$~nm compared to the FND PL outside the cavity (Fig~\ref{fig:scheme}c, black trace), which we will investigate in detail in the following.  

To fabricate the microcavities, we first deposit a fully reflective 100~nm Ag layer on Si \textit{via} electron beam evaporation physical vapor deposition to create flat, uniform substrates for spin coating.
For the FND cavities, $\sim$120~nm FNDs (see Supplementary Information (SI) for details) containing  1~ppm of NV centers were dispersed at a concentration of 0.5~mg/mL in a 1:3 water:1-propanol solution containing polyvinyl pyrrolidone (PVP) at 2.5~wt\%.
This suspension was spin-coated at speeds from 600 to 2600 rpm to produce FND-doped polymer thin-films ranging in thickness from 107--210~nm. 
Finally, a semi-transparent Ag layer was deposited \textit{via} electron beam evaporation to complete the microcavity.
For hBN, this fabrication protocol was repeated using as-received hBN particles (Graphene supermarket, 70~nm particle size) dispersed at a concentration of 0.3~mg/mL in a solution containing polymethyl methacrylate (PMMA) in chlorobenzene at 3.0~wt.\%. 
These particles are known to contain a range of photoluminescent spin-active defects that emit from the violet to near-infrared spectral region~\cite{Singh_AM2025}.

Fig.~\ref{fig:scheme}d shows a photograph of a $20\times20$~mm Si-wafer substrate, half of which is coated with a complete FND-doped microcavity (left), while the other half lacks the top Ag mirror for control experiments (right).  
The light absorption by the resonant cavity mode produces a green tint in the cavity region, which is not present in the no-cavity region. 
Fig.~\ref{fig:scheme}e shows a confocal PL image of the FND microcavity, showing evenly distributed FND particles throughout the field of view. 
For both FND and hBN microcavities, we fabricated a suite of devices with fundamental optical modes spanning the spectral region from 500--800~nm, covering the  spectral emission range of the NV center in FNDs and the hBN emitters. 
Supplementary Fig.~\ref{fig:reflectance}a~and~b show the normal-angle reflectance spectra for the FND and hBN microcavities, respectively, clearly showing reflectance anti-peaks with reflectivity decreases of 60--80\% at the predicted resonance mode energies (see Supplementary Fig.~\ref{fig:reflectance}).

\begin{figure*}[t!]
    \centering
    \includegraphics[width=\linewidth]{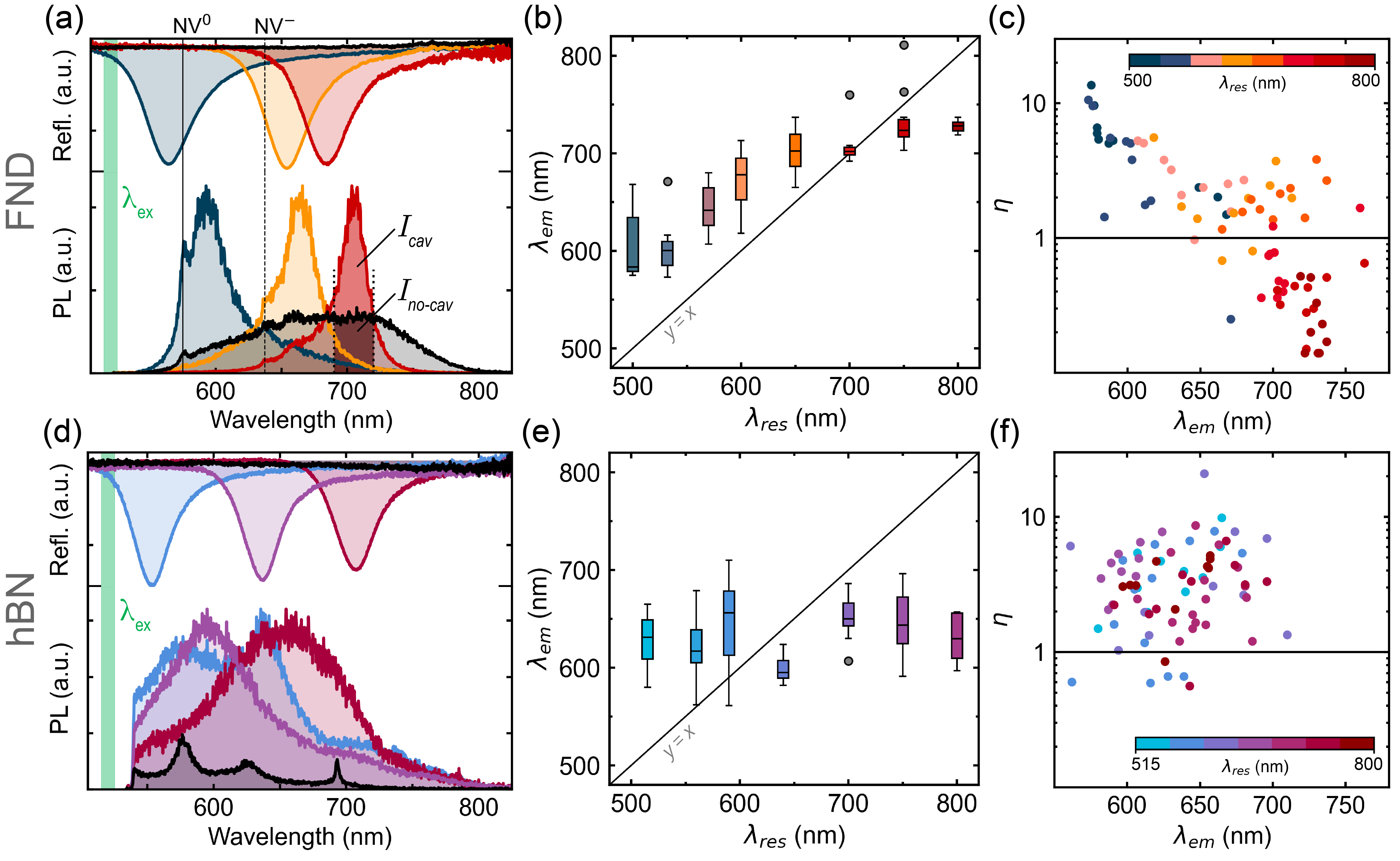}
    \caption{(a,d) Thin-film cavity reflectivity spectra (top, colored traces) and corresponding typical PL spectra of FND (a) and hBN NPs (d) quantum sensors embedded in the cavity (bottom, colored traces). Black traces at the top of a) and b) show the no-cavity reflectivity and no-cavity PL spectra of FNDs (a) and hBN (d) at the bottom. The green bar on the left indicates the excitation wavelength $\lambda_{ex}$= 520 nm. 
    (b,e) Box and whisker plot of  $\lambda_{em}$ as a function of $\lambda_{res}$ for all investigated FNDs (b) and hBN NPs (e). The horizontal dash in the box represents the average, the edges of the box the upper and lower quartiles, and the whiskers the minimum and maxiumum values of  $\lambda_{em}$. Statistical outliers (1.5 × interquartile range) are in grey. 
    (c,f) Scatter plot of cavity enhancement factor $\eta$ (see main text for details) as a function of  $\lambda_{em}$ for individual FNDs (c) and hBN particles (f) , where the color of each dot indicates  $\lambda_{res}$ of the cavity.}
    \label{fig:spectra}
\end{figure*}

First, we investigated the effect of the cavities on the FND and hBN NP spectral PL characteristics and PL brightness changes in the spectral region of the cavity resonance (Fig.~\ref{fig:spectra}). 
We used a confocal PL microscope with 5 ps pulsed 520~nm laser excitation and collected PL above 540~nm in all experiments (see Supplementary Information for details). 
We collected PL spectra and time-resloved PL decay traces for FNDs and hBN NPs inside and outside cavities, which are analyzed and discussed in detail in the following text and in Fig.~\ref{fig:spectra} and Fig.~\ref{fig:pl_data}.

Typical reflectance and PL spectra for three FND cavity devices are depicted in Fig.~\ref{fig:spectra}a, where the reflectance anti-peak, a signature of the resonant cavity mode energy, is color-coded to the corresponding PL spectrum from FNDs inside the cavity.
The 'no-cavity' control device showed no absorption in the reflectance profile (black trace), indicating the absence of a cavity resonance and the FND PL spectrum (Fig.~\ref{fig:spectra}a, black trace) shows a typical NV emission profile with contributions from the NV$^0$ and NV$^-$ charge states. 
The PL spectra of FNDs inside the cavity (Fig.~\ref{fig:spectra}a, colored traces) were strongly modulated and significantly narrower, with the PL peak position $\lambda_{em}$ shifting with the cavity resonance  peak position $\lambda_{res}$. 
The characteristic NV$^0$ and NV$^-$ zero phonon lines at 575~nm and 637~nm, respectively, are clearly visible in all spectra, confirming the NV center as the origin of the PL.

While each cavity resonance shown in Fig.~\ref{fig:spectra}a is representative of the entire thin-film microcativy, the PL spectra were acquired from individual FNDs. 
The spectra in Fig.~\ref{fig:spectra}a suggest that $\lambda_{em}$ is red shifted from $\lambda_{res}$ by 10-20~nm. 
To determine if this shift is typical for most FNDs in all thin-film cavities, we acquired PL spectra from 10 individual FNDs in each cavity. 
Fig.~\ref{fig:spectra}b shows a box and whisker plot of  $\lambda_{em}$ as a function of $\lambda_{res}$ for all investigated FNDs. 
The horizontal dash in each box represents the average, the edges of the box the upper and lower quartiles, and the whiskers the minimum and maximum values of $\lambda_{em}$. 
Statistical outliers (1.5 × interquartile range) are in gray. 
The average $\lambda_{em}$ linearly increases with $\lambda_{res}$ for $\lambda_{res}$ values between 525~nm and 650~nm and only slighty increases for $\lambda_{res}>650$~nm. 
In the $\lambda_{res}$ range of 500--650~nm,  $\lambda_{em}$ is on average red-shifted by more than 50~nm, with a maximum shift of 80~nm for  $\lambda_{res}=500$~nm. 
Above 700~nm, $\lambda_{em}$ is slightly blue-shifted relative to $\lambda_{res}$. 
Overall, Fig.~\ref{fig:spectra}b demonstrates that the average spectral emission peak position of FND NV PL can be reliably tuned by the cavity resonance. 

The discrepancy between $\lambda_{em}$ and $\lambda_{res}$ is likely caused by a local modification of the effective refractive index $n_{eff}$ of the polymer thin-film by the FND particles. The reflectance spectra in Fig.~\ref{fig:spectra}a represent averages over an 100 $\times$ 100 µm$^2$ area of the cavity and only a small fraction of this area contains FNDs. The PL spectra on the other hand are acquired using a focused laser beam with a diameter of ca. 300 nm, where an FND covers a significant fraction of the illuminated area. In this area and associated thin-film volume, the FND increases $n_{eff}$ of the thin-film due to the higher refractive index of diamond ($n$=2.4) compared to pure PVP ($n$=1.5), leading to a red-shift in the cavity resonance wavelength. 

We then investigated the FND PL intensity enhancement in the spectral region of the cavity resonance. 
We define this spectral cavity enhancement  
$\eta = {I_{cav}} / I^{avg}_{no-cav}$, for constant laser power impinging on the device,
where \textit{I$_{cav}$} and \textit{I$^{avg}_{no-cav}$} are the integrated PL intensities  $\pm15$~nm around $\lambda_{em}$ of individual FND PL spectra inside the cavity and the average PL spectrum of FNDs outside the cavity, respectively (see SI Fig.~\ref{fig:fnd_gauss} for details). 
Fig.~\ref{fig:spectra}c shows a scatter plot of $\eta$ as a function of  $\lambda_{em}$ for individual FNDs, where the color of each dot indicates  $\lambda_{res}$ of the cavity. 
We find the strongest enhancement of up to  $\eta = 10$ for the cavity with  $\lambda_{res} = 500$~nm and $\lambda_{em}$ emission peak positions between 575--600~nm, close to the ZPL of NV$^0$ at 575~nm. 
On average,  $\eta$  drops significantly with increasing $\lambda_{em}$ and $\eta$ varies significantly between individual particles with most enhancement values between 1-3 for $\lambda_{em}$ between 625--700~nm. 
Above $\lambda_{em} = 700$~nm, FNDs show $\eta$  values above and below 1, indicating that both PL enhancement and suppression occur in this region.

In principle, both the excitation beam ($\lambda_{ex} = 520$) and NV center PL can couple to the cavity resonance mode. Our experiments cover three cases. 1) The excitation beam and NV PL couple to the cavity resonance  ($\lambda_{res} = 500-570$~nm ), and we observe the strongest spectral emission enhancement of up to $\eta = 10$.  2) Mostly NV PL couples to the cavity resonance ($\lambda_{res} = 600-700$~nm). Here, the cavity enhancement is still present ($\eta > 1$) but reduced, since most excitation light is reflected (see red and orange traces in Fig.~\ref{fig:spectra}a). 3) NV PL only weakly couples to the cavity resonance ($\lambda_{res} > 700 nm$ ). The NV PL decreases above 700 nm and most excitation light is reflected resulting in suppression of NV PL inside the cavity compared to outside the cavity in most cases (see red markers in Fig.~\ref{fig:spectra}c), likely due to optical losses in the top mirror. 

We then performed the same analysis for hBN NPs. The hBN particles studied here contain a range of carbon-related defects~\cite{Singh_AM2025}, whose spectral PL characteristics vary greatly between individual particles, with emission from 550--800~nm upon 520~nm excitation (see Supplementary Fig.~\ref{fig:hbn-control} for examples). 
Fig.~\ref{fig:spectra}d shows the PL spectrum of an hBN particle outside the cavity (black trace) compared to spectra of hBN particles in three different cavities (colored traces). 
Unlike for FNDs, there is no systematic correlation between  $\lambda_{res}$ and  $\lambda_{em}$. 
Due to the great variation in hBN PL spectral characteristics, even the identification of a single emission peak position is non-trivial (see Supplementary Fig~\ref{fig:hbn-gauss} for details). 
Fig.~\ref{fig:spectra}b shows a box and whisker plot of  $\lambda_{em}$ as a function of $\lambda_{res}$ for all investigated hBN particles. 
It confirms that $\lambda_{em}$ of most hBN particles inside the cavity is between 600--650~nm, irrespective of $\lambda_{res}$, with a large spread of $\lambda_{em}$ values between 570~nm and 700~nm. 
Interestingly, we nonetheless observe that the PL of more than 90\% of hBN particles in cavities is enhanced 2-10 times compared to particles outside the cavity (Fig.~\ref{fig:spectra}f). The hBN particles generally appeared slightly larger than FNDs in confocal PL images (see Supplementary Fig.~\ref{fig:PLmaps}) suggesting hBN NP aggregation. Due to their larger size, the aggregates may have been in closer proximity to the top and bottom silver mirror, leading to surface plasmon-related enhancements that are independent of the cavity resonance.

\subsection{Emission Rate Enhancement}
\begin{figure*}[t!]
    \includegraphics[width=\linewidth]{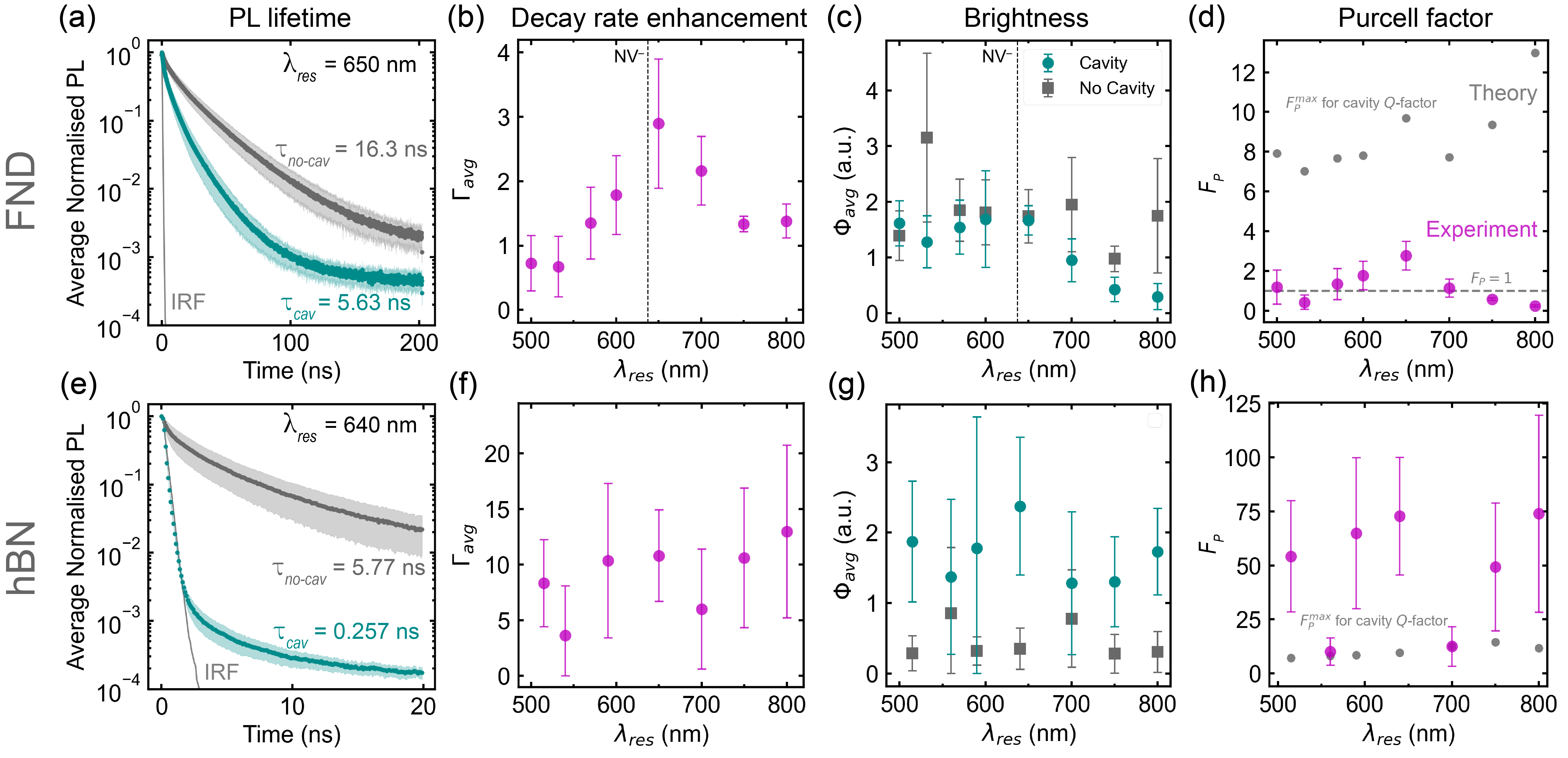}
    \caption{\textbf{Emission enhancement of FNDs and hBN NPs inside thin-film cavities---} (a,e) 
    Averaged time-resolved PL decay traces of FNDs (a) and hBN NPs (e) inside (green traces) and outside (grey traces) the cavities. 
    The shaded envelopes show the standard deviation between individual measurements and the black dotted line the instrument response function (IRF).
    (b,f) Average emission rate enhancement $\Gamma_{avg}$ for FND (b) and hBN (e) cavities as a function of cavity resonance wavelength $\lambda_{res}$. 
    (c,g) Average total PL brighness $\Phi$ for FND (c) and hBN (f) cavities as a function of cavity resonance wavelength $\lambda_{res}$. 
    (d,h) Experimental (pink markers) and theoretical (grey markers) Purcell enhancement factor $P_F$ for FND (d) and hBN (h) cavities as a function of cavity resonance wavelength $\lambda_{res}$.}
    \label{fig:pl_data}
\end{figure*}

Having established the effect of the cavity on the quantum sensors' spectral PL properties, we investigated the effect of the cavities on the PL lifetime and total PL brightness to quantify the cavity-induced modulation of FND and hBN PL.
Fig.~\ref{fig:pl_data}a shows the average time-resolved PL decay traces of ten FNDs inside and 10 FNDs outside the cavity with  $\lambda_{res} = 650$~nm. 
The shaded envelope represents the standard deviation between individual measurements.
FNDs inside the cavity show a significantly shorter PL lifetime, of  $\tau_{cav} = 5.63$~ns compared to $\tau_{no-cav} = 16.3$~ns for FNDs outside the cavity, determined by an amplitude-weighted average of a double exponential fit to the decays. 
To quantify the cavity-induced reduction in the PL lifetime we investigated the decay rate enhancement  $\Gamma$, which we define as the ratio of the observed decay rate $\gamma$ inside and outside the cavity.
\begin{equation}
\Gamma=\frac{\gamma_{cav}}{\gamma^{avg}_{no-cav}}, 
\end{equation}
where the decay rate $\gamma$ = $\tau^{-1}$ and $\gamma_{cav}$ and $\gamma^{avg}_{no-cav}$ are the decay rate of ten individual FNDs inside the cavity and the average decay rate of ten FNDs outside the cavity, respectively.
Fig.~\ref{fig:pl_data}b shows the average decay rate enhancement  $\Gamma_{avg}$ of ten investigated FND cavities per cavity device as a function of the cavities' spectral resonance peak position  $\lambda_{res}$.
Error bars represent the standard deviation between individual measurements. 
Cavities with $\lambda_{res}$ between 575~nm and 800~nm exhibit a $\Gamma_{avg} > 1$, with a peak enhancement of 2.9 for $\lambda_{res} = 650$~nm. 
Only cavities with resonances below 600~nm show no statistically significant change in the decay rate. 
Hence, all cavity devices with resonances in the spectral region where the NV emits, \textit{i.e.} 570--800~nm, show a decay rate enhancement. The fact that we observe strong spectral PL enhancement $\eta$ for cavities with resonances below 600~nm (Fig.~\ref{fig:spectra}c), where no emission rate enhancement is present (Fig.~\ref{fig:pl_data}b) suggests that cavity enhancement of the excitation beam causes the spectral PL enhancement in this spectral region.

The measured PL decay rate $\gamma$ is the sum of radiative $\gamma^{rad}$ and non-radiative decay $\gamma^{nr}$ rates.
To identify whether the observed increase in $\gamma$  is the result of an increase in $\gamma^{rad}$,   $\gamma^{nr}$ or both, we investigated the total PL brightness $\Phi$ of FNDs, which is proportional to the PL quantum yield $\gamma^{rad}$/$\gamma^{rad}+\gamma^{nr}$. 
We determined $\Phi$ for FNDs inside and outside the cavity as the total area under the PL spectra for each FND.  
Fig.~\ref{fig:pl_data}c shows \textit{$\Phi_{avg}$} for all devices as a function of $\lambda_{res}$ for FNDs inside and outside the cavity. 
Markers represent the brightness averaged over 10 FNDs per sample and the error bars the standard deviation between individual FNDs. 
In half of the devices, FNDs inside and outside the cavity exhibit a similar brightness, with some devices showing a roughly 2-fold decrease ($\lambda_{res}=525$~nm and $\lambda_{res}=700$~nm and above). 
Importantly, the FND brightness is the same inside and outside the cavity (within the standard deviation) for three devices with $\lambda_{res}$ values of 575--650~nm, for which we see significant decay rate enhancements. 
Given the observed NV decay rate enhancement of 2.9 observed for $\lambda_{res}=650$~nm and assuming no optical losses in the Ag top mirror and an NV quantum yield of 0.5~\cite{Plakhotnik_DaRM2018}, this would suggest that both $\gamma^{rad}$ and $\gamma^{nr}$ increase by a factor of 2.9, which would lead to no change in the NV quantum yield and brightness. 
However, considering light absorption in the Ag mirror is present but difficult to quantify in our cavities, an increase in the radiative rate is likely dominant. 

To investigate whether the observed enhancement can be understood in terms of a Purcell enhancement, we determined an experimental Purcell enhancement factor $F_P$~\cite{Englund_NL2010} and compared it to a theoretical enhancement factor $F_P^{th.}$. 
These are defined as
\begin{equation}
    F_P = \frac{\Phi_{cav}}{\Phi_{no-cav}}\frac{\gamma_{cav}}{\gamma_{no-cav}}
    \label{eq:purcell}
\end{equation}
where $\Phi_{cav}$ and  $\Phi_{no-cav}$ are the average PL brightness inside and outside the cavity, respectively, (Fig.~\ref{fig:pl_data}c) and $\gamma_{cav}$ and $\gamma_{no-cav}$ are the observed PL decay rates inside and outside the cavity, respectively. 
The theoretical Purcell factor can be estimated for these devices as
\begin{equation}
    \label{}
    F_{P}^{th.} \simeq \frac{6Q}{\pi^2}
\end{equation}
where $Q$ is the cavity quality factor (see Supplementary Information for further details).

Fig.~\ref{fig:pl_data}e shows the $F_{P}$ and $F_{P}^{th.}$ for all FND cavities as a function of $\lambda_{res}$. 
Most cavities exhibit a theoretical Purcell enhancement of $F_{P}^{th.} = 7.5 \pm 1.0$. 
The cavity with $\lambda_{res}= 650$~nm has a slightly higher $F_{P}^{th.}$ of 9.7, coinciding with the highest observed experimental $F_{P}$ of 2.9 in the spectral region of the NV$^-$ ZPL. 
In the spectral region above 700~nm where the NV PL begins to drop off, $F_{P}^{th.}$ slightly increases while $F_{P}$ decreases. 
In all cases $F_{P}$ is significantly lower than $F_{P}^{th.}$, which is expected since cavity losses, \textit{e.g.} in the Ag mirrors, are not considered in our model. 
However, the three devices with resonances from 575--650~nm, for which the highest decay rate enhancement was observed, all show average $F_{P} > 1$, suggesting that a Purcell enhancement of the NV radiative rate causes the observed enhancements.  

\subsection{hBN Emission Rate Enhancement}

We now focus on the hBN cavities, the results of which are presented in the bottom row of Fig.~\ref{fig:pl_data}. 
Unless noted otherwise, the hBN data were analyzed as described for FNDs above.  
Fig.~\ref{fig:pl_data}e shows the average time-resolved PL decay traces of hBN particles inside and outside the  $\lambda_{res} = 640$~nm cavity up to 20 ns after the excitation pulse, \textit{i.e.} on a timescale ten times shorter than for FNDs in  Fig.~\ref{fig:pl_data}a. 
While hBN particles outside the cavity show an average PL lifetime of $\tau_{no-cav} = 5.77$~ns (amplitude-weighted average of a double exponential fit), $>99$\% of photons in the cavity are emitted within the first 2 ns of the decay and faster than our instrument response function, suggesting a dominant PL lifetime  $\tau_{cav} < 0.5$~ns.
Only a very small fraction ($<0.1$\%) of photons are emitted after 3 ns, with a lifetime of 6.82~ns, similar to $\tau_{no-cav}$.

Fig.~\ref{fig:pl_data}f shows the resulting decay rate enhancement $\Gamma_{avg}$ for all hBN devices as a function of $\lambda_{res}$. 
We find significant decay rate enhancements for all devices with average values for different devices ranging from 4 to 13 and also strong variations between individual particles within the same device. 
Fig.~\ref{fig:pl_data}f  shows the corresponding average hBN brightness as a function of  $\lambda_{res}$. 
Unlike for FNDs, we find a systematic increase in PL brightness for hBN particles inside the cavity compared to outside and no systematic dependence on $\lambda_{res}$. 
We also observed that the hBN NPs inside the cavity showed pronounced fluctuations in PL intensity over time.
These fluctuations were also observed in real-time in steady-state widefield measurements and is a behavior not observed in the no-cavity control devices. 
This blinking may arise from transient switching of defect centers between radiative and non-radiative states~\cite{White_PRA2020}.
Coupling to the planar metallic microcavity may enhance these effects through resonant interactions, including \textit{via} Purcell effects and plasmonic coupling, where fluctuations in the charge state of the emitter may be amplified in the cavity, leading to the temporally unstable blinking observed in these measurements.
These observations indicate that the cavity not only modifies spectral features but also reveals dynamic emission behaviors that are absent in uncoupled nanoparticles.

Fig.~\ref{fig:pl_data}e shows the Purcell enhancement factors $F_{P}$ and $F_{P}^{th.}$ as a function of cavity $\lambda_{res}$. 
Due to the combination of a strong PL decay rate enhancement and PL brightness enhancement, we obtain $F_{P}$ values of up to 75, which is an order of magnitude higher than the highest theoretical enhancement $F_{P}^{th.}$. 
We therefore conclude that Purcell enhancement alone cannot explain the observed extreme modulation of hBN PL in our cavities. 

Several mechanisms may explain our observations. Both the excitation beam and the hBN PL can be enhanced by cavity resonances. Since we don't observe a clear dependence of emission enhancement on $\lambda_{res}$, it is unlikely that cavity resonances are the main cause for the enhancement. The close proximity of the hBN particles to the top and bottom silver mirror may cause surface plasmon-related enhancements. Direct electrical contact  between hBN particles and one of the mirrors can also lead to a fast light-induced redistribution of charges that may explain the fast PL decay. Lastly, a local change in cavity geometry due to the formation of hBN aggregates could alter the cavity properties around particles and may focus light onto hBN particles independent of the polymer layer thickness. Future studies will investigate the origin of the observed PL modulation. 


\subsection{Enhancement of Quantum Sensing}

Finally, we investigated whether the observed enhancement of FND and hBN PL inside cavities improves their quantum sensing performance. 
We first focused on FNDs. We employed a custom built wide-field quantum sensing microscope~\cite{Scholten_JAP2021} (Supplementary Fig.~\ref{fig:SI_widefield} to acquire ODMR spectra for FNDs inside the cavities investigated in the previous section as a function of cavity resonance $\lambda_{res}$. 
See Supplementary Information for full experimental details. 
These devices were fabricated on Si wafer substrates, which strongly absorb microwaves. 
Hence, microwaves were delivered \textit{via} a loop antenna from the top-mirror side of the cavity, which made a direct comparison between FNDs inside and outside the cavity challenging due to top Ag mirror MW absorption only being present on the cavity side. 
However, it allowed us to investigate the ODMR contrast for all cavity resonances. 
We find the highest ODMR contrast of 4.4\% for the cavity with $\lambda_{res}= 650$~nm, which gradually drops to approximately 2\% for resonance wavelengths of 500~nm and 800~nm (see Supplementary Fig.~\ref{fig:SI_contrast_res}).

We were unable to acquire ODMR spectra for the hBN thin-film cavities. 
In general, the positive ODMR contrast of the spin pair emitters in the hBN NPs investigated here is below 1\% even under optimized MW delivery conditions~\cite{Singh_AM2025,Scholten_NC2024}. 
Hence, one possible explanation is that the MW field inside our cavities was too low to efficiently drive the hBN spin transitions. 
For the hBN emitters investigated here, the MW-induced spin mixing that is the basis of ODMR occurs in a metastable state~\cite{Scholten_NC2024}.  Hence, another possible explanation is that the PL decay observed for most hBN NPs inside the cavity became so fast, i.e. the excited state lifetime so short, that transitions to the metastable state were very inefficient resulting in a negligible population of the metastable state. 
Lastly, the decrease in hBN PL stability over time increased the measurement noise in ODMR spectra.   

\begin{figure}[H]
\includegraphics[width=0.5\linewidth]{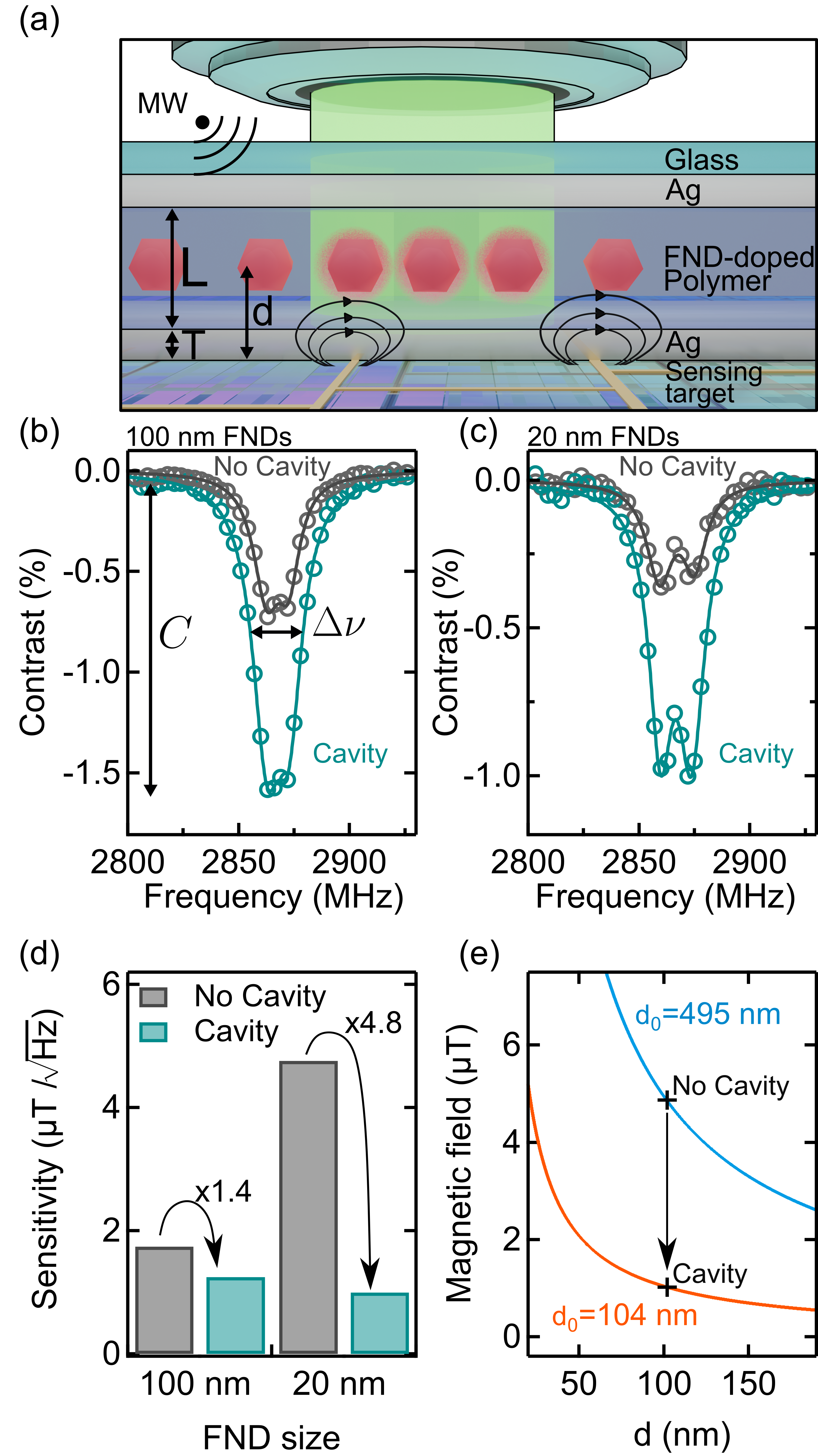}
    \caption{\textbf{Cavity-enhanced magnetic sensing---} 
    (a) Schematic illustration of magnetic imaging of a microelectronic circuit using FNDs in a thin-film cavity. 
    (b-c)  Average ODMR spectra of 100~nm (b) and 20~nm (c) FNDs inside and outside the cavity. 
    (d) Magnetic field sensitivity for 100~nm and 20~nm FNDs inside and outside the cavity. 
    (e) Magnetic field B at the location of the nanoscale sensor as a function of its distance $d$ from a source (e.g. an electrical current in a microchip) for our cavity (orange trace) and a hypothetical thicker cavity (blue trace). 
    The black markers indicate the smallest B-field detectable per $\sqrt{Hz}$  for 20 FNDs inside and outside the cavity.
    }  
    \label{fig:sensing}
\end{figure}

Based on these results, we focused on the FND cavities, selected the $\lambda_{res}= 650$~nm cavity for further analysis and fabricated devices on quartz substrates that allow MW delivery and imaging through the substrate (Fig.~\ref{fig:sensing}a) and hence a direct comparison between FNDs inside and outside the cavity. 
We fabricated one device with 100~nm FNDs and one with 20~nm FNDs (Adamas Nanotechnologies, USA) to also investigate if the much dimmer PL, mostly from the NV$^0$ charge state of the 20~nm FNDs \cite{wilsonEffectParticleSize2019} can be enhanced for ODMR-based quantum sensing.
Reflectance spectra for these devices is shown in Supplementary Fig.~\ref{fig:fnd_q_reflectance}.
Figs.~\ref{fig:sensing}b and c show ODMR spectra for 100~nm and 20~nm FNDs, respectively, inside and outside the cavity, with the characteristic NV$^-$ resonance at 2.87~GHz. 
The ODMR contrast increases by a factor of 2.2 (100~nm FNDs) and 2.8 (20~nm FNDs) while the full-width half-maximum (FWHM), indicated by the black arrows, remains constant within $\pm1$~MHz.  
The PL brightness $\Phi$ increases 3-fold for 20~nm FNDs inside the cavity, while it decreases by about half for the 100~nm FNDs.
In both cases, however,  we observe spectrally narrower PL in the NV$^-$ spectral region between 650 and 700 nm (20 nm FNDs) and 700 and 800 nm (100 nm FNDs) for FNDs inside the cavity (see Supplementary Fig.~\ref{fig:SI_spectra}). 
Spectral narrowing likely plays a dominant role in the enhancement of ODMR contrast observed in both samples’ cavity regions and suggests that the NV coupling to the cavity can selectively enhance NV$^-$ in our cavities.

Based on these measurements, we calculated the FND magnetic field sensitivity using~\cite{rondin_magnetometry_2014} 
\begin{equation}
S = \frac{4}{3\sqrt{3}}\frac{h}{g_e\mu_B}\frac{\Delta\nu}{C\sqrt{R}}
\end{equation}
where $h$ is Planck's constant, $g_e\sim$ 2.003 is the NV$^-$ centers electronic g-factor, $\mu_B$ is the Bohr magneton, $\Delta\nu$ is the FWHM of the ODMR spectrum determined via a double Lorentzian fit, $C$ is the ODMR contrast, and $R$ the PL photon count rate. 
Fig.~\ref{fig:sensing}d shows $S$ for 100~nm and 20~nm FNDs inside and outside the cavity. 
It reveals a 1.4-fold improvement in sensitivity for 100~nm FNDs and a 4.8-fold improvement for 20~nm FNDs. 
The 4.8-fold improvement for the 20~nm FNDs is driven by a combination of increase in ODMR contrast and PL brightness. 

To investigate the trade-off between cavity enhancement and sensor proximity to the sensing target, we estimated the magnetic field $B$ at the location of the FND sensor as a function of its distance from a source (\textit{e.g.} an electrical current in a microchip). 
Fig.~\ref{fig:sensing}e shows $B(d) = \frac{S}{d/d_0}$ , where $S = 1~\mu\text{T}$  (the smallest magnetic field the cavity-enhanced 20~nm FNDs can detect per $\sqrt{Hz}$ ) and $d$ is the distance between sensor and sensing target. 
The orange trace represents $d_0$ = $L/2+T = 104$~nm, which is the distance between 20~nm FNDs and the sensing target in our devices. 
The markers indicate the smallest B-field detectable per $\sqrt{Hz}$  for 20~nm FNDs inside and outside the cavity in our devices, respectively. 
For a hypothetical thicker cavity with $d_0 = 495$~nm (blue trace in Fig.~\ref{fig:sensing}e) and assuming the same 4.8-fold enhancement observed for our devices, the cavity enhancement would not offer any benefits over the thinner no-cavity devices investigated here. 
Hypothetical thicker cavities will need create stronger enhancements and have lower optical losses to compensate for the larger separation between FND sensor and the sensing target. 
This trade-off between cavity thickness and cavity enhancement will guide the design of future thin-film cavity devices using low-loss cavity materials. 


\section{Conclusions}
We have demonstrated cavity-mediated enhancement of emission in a suite of cavity devices containing fluorescent nanodiamonds and hexagonal boron nitride nanoparticles, underscoring the efficacy of Fabry-P\'erot microcavities in enhancing the fluorescence properties of this class of quantum sensors.
In the case of FND, this emission enhancement was captured by a Purcell enhancement model.
These observed enhancements in emission rates and ODMR signal contrast with FND suggests significant potential for these systems in quantum sensing applications, allowing for wafer-scale fabrication of enhanced quantum sensors for optimized device integration.

\begin{acknowledgement}
This work was supported by the Australian Research Council (ARC) through grants DE200100279, DP220102518, DP250100125, DP230101764, DE230100192, FT200100073. 
P.R. acknowledges support through RMIT University Vice-Chancellor’s Senior Research Fellowship. 
This work was supported by the ARC Centre of Excellence in Nanoscale Biophotonics (CNBP) (CE140100003).
This research was in part carried out at the RMIT Micro Nano Research Facility (MNRF) in the Victorian Node of the Australian National Fabrication Facility (ANFF-Vic) and the RMIT Microscopy and Microanalysis Facility (RMMF).
\end{acknowledgement}


\newpage
\setcounter{equation}{0}
\setcounter{figure}{0}
\setcounter{table}{0}
\setcounter{page}{1}
\setcounter{section}{1}
\renewcommand{\theequation}{S\arabic{equation}}
\renewcommand{\thefigure}{S\arabic{figure}}
\renewcommand{\thetable}{S\arabic{table}}

\begingroup
\centering
\section{Supplementary Information}
\includegraphics[width=\linewidth]{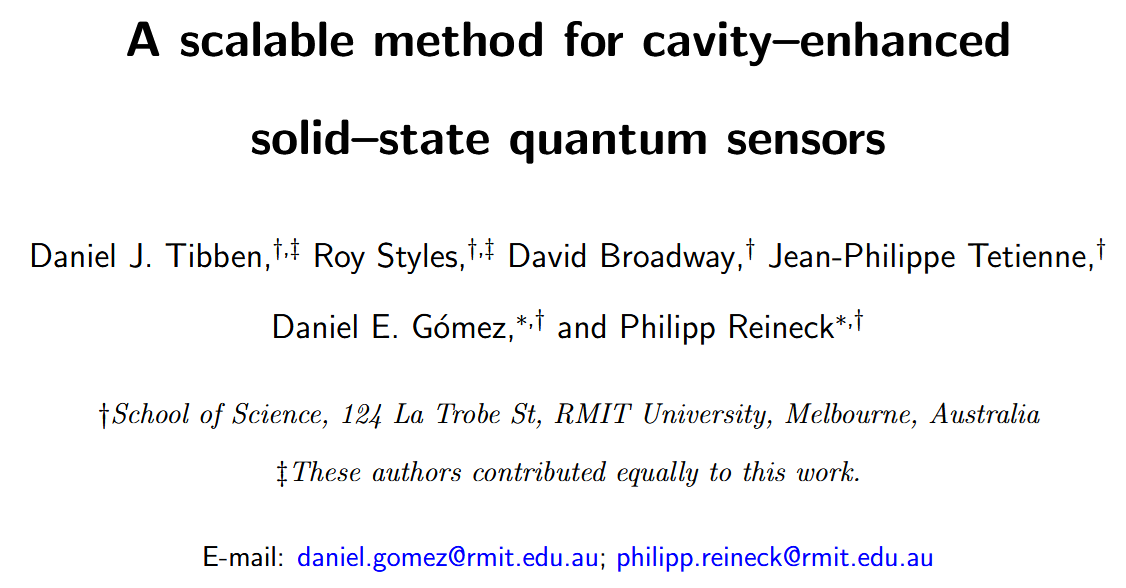}
\endgroup

\newpage

\subsection{Fluorescent Nanodiamonds} \label{sec:FND_description}

FNDs were fabricated by irradiating HPHT nanodiamonds (Nabond, China) with 2 MeV electrons to a fluence of 1 × 10$^{18}$ electrons cm$^{-2}$, then annealing them in vacuum at 900 °C for 2 hours, and oxidizing the particle powder in air at 520 °C for 2.5 h.

\subsection{Normal Angle Reflectance of FND and hBN NP Microcavity Devices}
\begin{figure}[H]
    \centering
    \includegraphics[width=\linewidth]{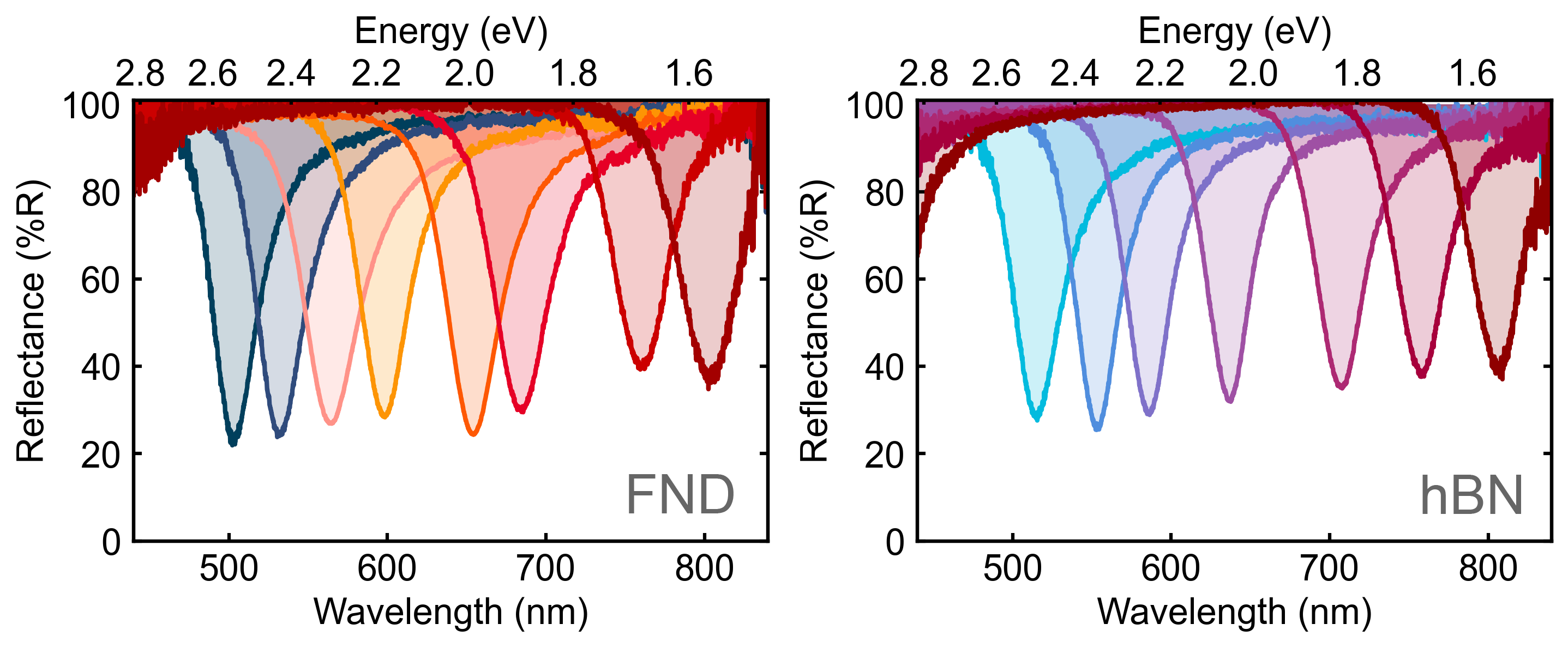}
    \caption{Normal angle reflectance of the cavity devices.}
    \label{fig:reflectance}
\end{figure}

The normal angle reflectance spectra, displayed in Fig.~\ref{fig:reflectance}, were used to determine the experimental resonance wavelengeth $\lambda_{res}$ using Eq.~\ref{eq:cav_approx} for the FND and hBN suites of devices.

\subsection{Transfer Matrix Model for Fabry--Pérot Microcavities}

The resonance energies of the Fabry--Pérot microcavity were calculated using a transfer matrix model. 
In this approach, each optical layer of thickness $d_i$ and refractive index $n_i(\omega)$ is represented by a propagation matrix, while interfaces between layers are described by Fresnel reflection and transmission coefficients. 
Multiplying these matrices yields the overall transfer matrix of the cavity. 
The reflectance spectrum is then obtained from the reflection amplitude $r(\omega)$, with $R(\omega) = |r(\omega)|^2$.  

The transfer matrix for a single layer of refractive index $n_i$, thickness $d_i$, and wavevector component 
\begin{equation}
    k_i = \frac{2\pi}{\lambda} n_i \cos\theta_i
\end{equation}
(at internal angle $\theta_i$) can be written as
\begin{equation}
    M_i =
    \begin{bmatrix}
    \cos(k_i d_i) & \tfrac{i}{q_i}\sin(k_i d_i) \\
    i q_i \sin(k_i d_i) & \cos(k_i d_i)
    \end{bmatrix},
\end{equation}
where $q_i = n_i \cos\theta_i$ for transverse electric (TE) polarization or $q_i = \tfrac{\cos\theta_i}{n_i}$ for transverse magnetic (TM) polarisation. The total transfer matrix for the full cavity is then
\begin{equation}\label{eq:tmm}
    M = \prod_i M_i,  
\end{equation}

with boundary conditions applied at the air/cavity and cavity/substrate interfaces.  

The Fabry--Pérot resonance condition corresponds to the round-trip phase in the cavity spacer, including mirror phase contributions, which can be simply modeled using
\begin{equation}
    2 n_{\text{eff}} L \frac{2\pi}{\lambda} + \phi_r = 2 m \pi,
\end{equation}
where $L$ is the polymer thickness, $n_{\text{eff}}$ is its effective refractive index, $\phi_r$ is the cumulative phase shift upon reflection from the silver mirrors, and $m$ is an integer mode index. 
Resonance energies $E = hc/\lambda$ therefore appear as minima in the reflectance spectrum, which can be directly compared with experimental reflectance dips. 
In unpolarized experiments, such as that in Fig.~\ref{fig:reflectance}, the resonance wavelength $\lambda_{res}$ is simply determined by 
\begin{equation}\label{eq:cav_approx}
    \lambda_{res} = 2n_{\text{eff}}L/m.
\end{equation}  
For FND-doped films, polyvinyl pyrrolidone (PVP) was used, and for hBN-doped films, polymethyl methacrylate (PMMA) was used.
Including the contributions from the top and bottom silver mirrors (and any contributions from the quantum sensor particles) the $n_{eff}$ was determined to be 2.04 for PVP cavities and 1.98 for PMMA cavities.

\begin{table}[H]
    \centering
    \renewcommand{\arraystretch}{1.2}
    \begin{tabular}{cc}
        \hline
        Device Type & Predicted Resonance $\lambda_0$ (nm)  \\
        \hline
        \hline
        PVP (FND-doped) & 500 \\
        PVP (FND-doped) & 532 \\
        PVP (FND-doped) & 570 \\
        PVP (FND-doped) & 600 \\
        PVP (FND-doped) & 650 \\
        PVP (FND-doped) & 700 \\
        PVP (FND-doped) & 750 \\
        PVP (FND-doped) & 800 \\
        \hline
        \hline
        PMMA (hBN-doped) & 515 \\
        PMMA (hBN-doped) & 560 \\
        PMMA (hBN-doped) & 590 \\
        PMMA (hBN-doped) & 640 \\
        PMMA (hBN-doped) & 700 \\
        PMMA (hBN-doped) & 750 \\
        PMMA (hBN-doped) & 800 \\
        \hline
        \hline
    \end{tabular}
    \caption{Modeled resonance wavelength of cavity devices fabricated on Si.}
\end{table}

\subsection{Purcell Enhancement in Fabry-P\'erot Micocavities}
To characterise the emission enhancement by the resonant conditions in the cavity, we describe our results in terms of the Purcell enhancement of both emission rates and brightness between the `cavity' and `no-cavity' devices.

The brightness $\Phi$ of our devices is defined as the number of photons emitted from the device across the collected spectral region. 
This is calculated by taking the integral of the spectral profile using the trapezoidal method:
\begin{equation}
    \label{eq:trapz_fnd}
		\Phi = \int_{x_a}^{x_b} f(x) dx \approx \frac{x_{b} - x_{a}}{2N} \sum_{n=1}^{N}(f(x)+f(x_{n+1}))
\end{equation}
where $x_a$ and $x_b$ are the start and end wavelength of the spectral region.

The Purcell model we use to characterize the enhancement in our devices was developed by \citet{Englund_NL2010}, and is described in Eq.~\ref{eq:purcell}.

The theoretical Purcell factor can be calculated for these devices by
\begin{equation}
  \label{eq:purcell_theoretical}
   F_P^{th.} = \frac{3 \lambda_{res}^3 Q}{4 \pi^2 n^3 V_{eff}}
\end{equation}
where $\lambda_{res}$ is the cavity resonance wavelength, $n$ is the refractive index of the cavity medium and $V_{eff}$ is the effective mode volume in the cavity. In this case, we use an idealized approach and assume perfect longitudinal confinement in the planar cavity, with no lateral confinement,  where the assumption $V_{eff} \simeq 0.125 (\frac{\lambda}{n})^3$ is appropriate~\cite{Gerard_SQDFAaNC2003}.

This reduces the theoretical Purcell factor calculation to 
\begin{equation}
   \label{}
   F_P^0 \simeq \frac{6Q}{\pi^2}
\end{equation}
which is a valid assumption for this class of Fabry-P\'erot microcavity device, where the lateral physical dimension exceeds the longitudinal physical dimension by at least 5 orders of magnitude, resulting in dominated confinement in the longitudinal direction and negligible confinement in the lateral direction. 

\subsection{Cavity $Q$-factor Determination}
\label{sec:q-factor}
The quality factor of the cavity devices was determined experimentally from the inverse reflectance spectra.

The cavity quality factor, $Q$, is defined as
\begin{equation}
    Q = \frac{\lambda_0}{\Delta \lambda}
\end{equation}
where $\lambda_0$ is the resonance wavelength and $\Delta \lambda$ is the full width at half maximum (FWHM) of the inverse reflectance peak. 
In the experimental reflectance spectra, $\lambda_0$ was identified as the wavelength at maximum inverse reflectance. The FWHM, $\Delta \lambda$, was determined from the wavelengths at which the inverse reflectance crossed half of the peak amplitude relative to a baseline. The baseline was calculated from the edges of the spectrum, and cubic interpolation was applied to the inverse reflectance data to determine the half-maximum crossing points accurately.

\begin{table}[H]
    \centering
    \small
    \begin{tabular}{ccc | ccc}
        \multicolumn{3}{c}{\textbf{FND}} & \multicolumn{3}{c}{\textbf{hBN}} \\
        \hline
        Resonance (nm) & \textit{Q}-factor & Purcell factor $F_P^{th.}$ & Resonance (nm) & \textit{Q}-factor & Purcell factor $F_P^{th.}$ \\
        \hline \hline
        500 & 13.02 & 7.92 & 515 & 11.76 & 7.15 \\
        532 & 11.57 & 7.03 & 560 & 13.48 & 8.19 \\
        570 & 12.62 & 7.67 & 590 & 14.06 & 8.55 \\
        600 & 12.85 & 7.81 & 640 & 15.72 & 9.56 \\ 
        650 & 15.95 & 9.70 & 700 & 20.66 & 12.56 \\
        700 & 12.72 & 7.73 & 750 & 23.82 & 14.48 \\
        750 & 15.39 & 9.36 & 800 & 19.21 & 11.68 \\
        800 & 21.34 & 12.97 &  &  &  \\
        \hline \hline
        Median Value: & 12.94 & 5.16 & Median Value: & 15.72 & 6.27 \\
        \hline
    \end{tabular}
    \caption{List of theoretical maximum Purcell factors for fluorescent nanodiamond and hBN cavity devices based on cavity Q-factor
    .   }
    \label{tab:purcell_theoretical}
\end{table}

\subsection{Cavity Enhancement Within a Gaussian Projection of the Cavity Mode}
\label{sec:cavity-enhancement-gaussian}
To extract a projection of the cavity mode on emission from the cavity, cavity emission spectra were analyzed by fitting a sum of two Gaussian functions,
\begin{equation}
I(\lambda) = A_1 \exp\!\left[-\frac{(\lambda - \mu_1)^2}{2\sigma_1^2}\right] + A_2 \exp\!\left[-\frac{(\lambda - \mu_2)^2}{2\sigma_2^2}\right],
\end{equation}
using nonlinear least--squares optimization. 
The cavity peak maximum was identified from the fit, and an integration range of $\pm 15$~nm was applied around this resonance.  

The integrated cavity emission in this range was compared with the corresponding integrated intensity from reference (no cavity) spectra. 
No cavity reference areas were averaged across multiple particles for fair comparison. The cavity enhancement ratio $\eta$ was defined as
\begin{equation}\label{eq:eta}
\eta = \frac{\int_{\lambda_{\mathrm{max}}-15}^{\lambda_{\mathrm{max}}+15} I_{\mathrm{cav}}(\lambda)\, d\lambda}{\big\langle \int_{\lambda_{\mathrm{max}}-15}^{\lambda_{\mathrm{max}}+15} I_{\mathrm{no-cav}}(\lambda)\, d\lambda \big\rangle}.
\end{equation}

\begin{figure}[H]
    \centering
    \includegraphics[width=\linewidth]{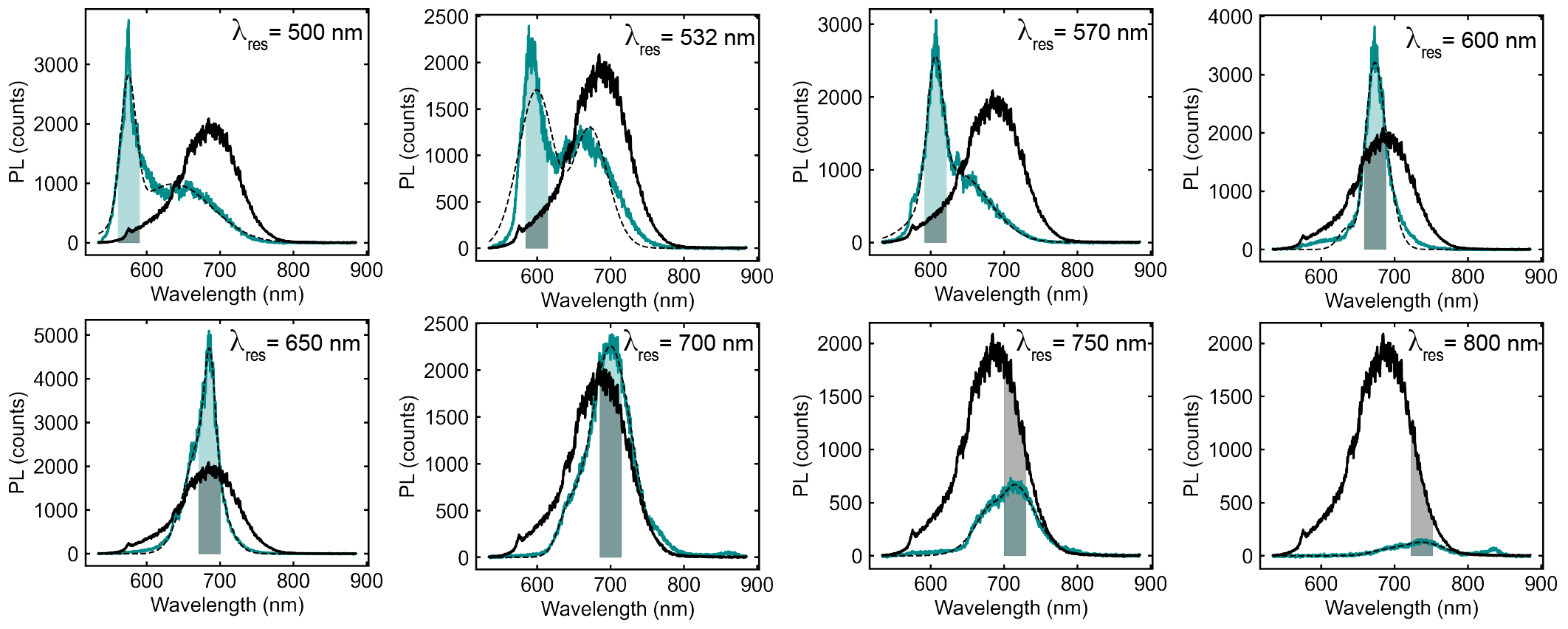}
    \caption{Fitting a double Gaussian projection (black dashed lines) of the cavity mode on example FND cavity emission spectra (teal line) and the average FND no cavity control emission spectrum (black line), as defined in Eq.~\ref{eq:eta}.}
    \label{fig:fnd_gauss}
\end{figure}

\begin{figure}[H]
    \centering
    \includegraphics[width=\linewidth]{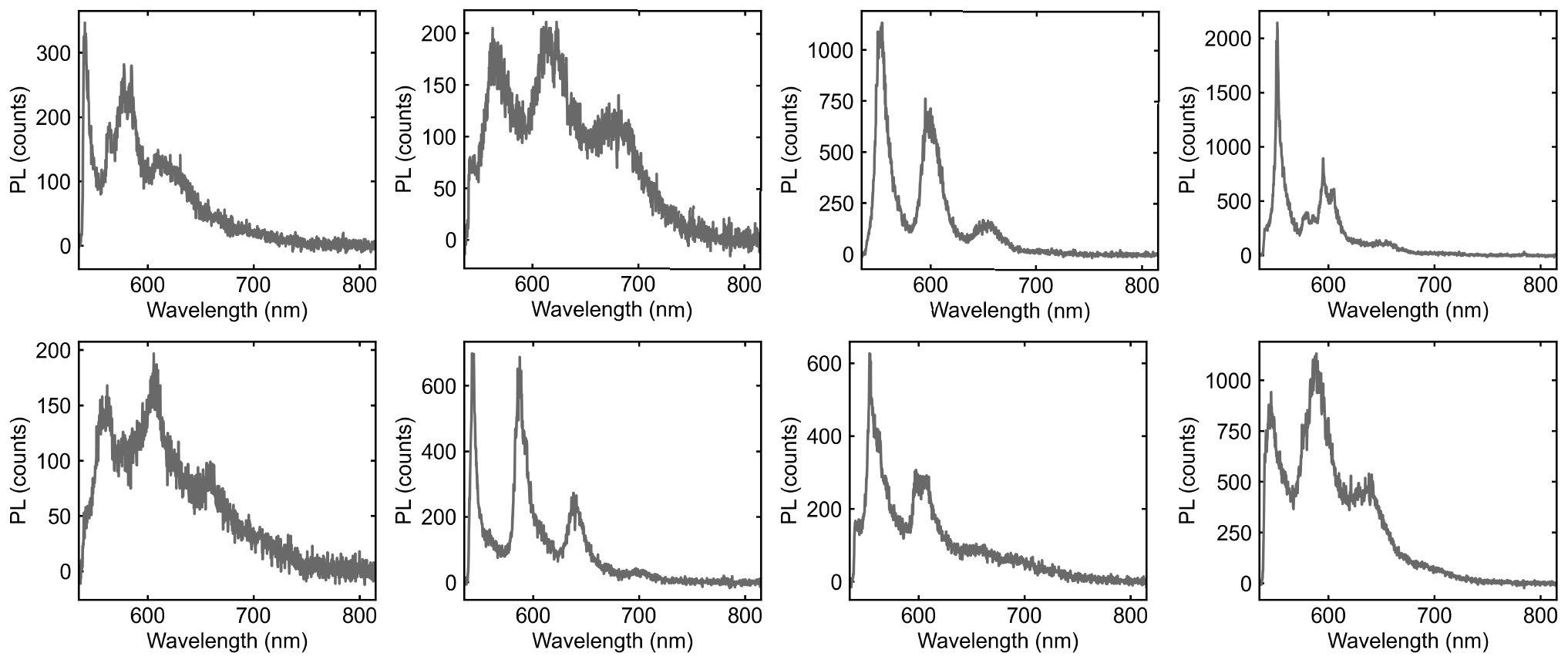}
    \caption{Photoluminescence variability in hBN particles outside the cavity.}
    \label{fig:hbn-control}
\end{figure}

\begin{figure}[H]
    \centering
    \includegraphics[width=\linewidth]{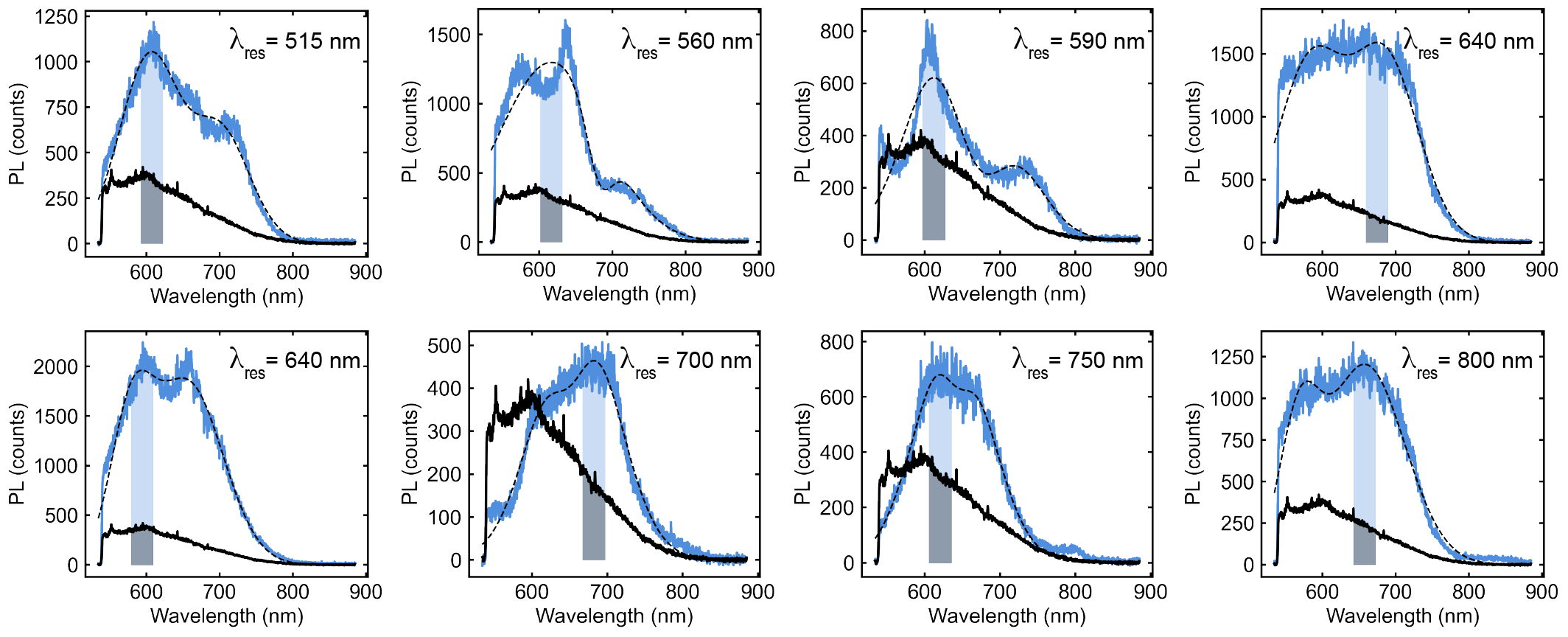}
    \caption{Fitting a double Gaussian projection (black dashed lines) of the cavity mode on example hBN cavity emission spectra (blue line) and the average hBN no cavity control emission spectrum (black line), as defined in Eq.~\ref{eq:eta}. 
    Emission maxima and lineshape of hBN cavity devices do not correlate well with the resonance energies determined from measured reflectance in Fig.~\ref{fig:reflectance}.}
    \label{fig:hbn-gauss}
\end{figure}

\subsection{hBN Emission Characteristics}

The PL spectra of hBN nanoparticles are usually non-uniform due to the heterogeneous distribution of luminescent centers, and their surrounding environments. 
Emission typically originates from intrinsic point defects such as vacancies, donor–acceptor pairs, and impurity complexes, each associated with distinct ZPLs and phonon sidebands. 
Theoretical predictions~\cite{Sajid_PRB2018,Auburger_PRB2021} and large-scale statistical studies~\cite{Islam_AN2024} have shown that hBN emitters can be grouped into several discrete spectral families within the $\sim1.6-2.2$~eV range, indicating the coexistence of multiple defect types even within nominally identical samples. 
Additionally, local variations in dielectric screening, strain, and surface termination can lead to shifts in emission energies and differences in spectral line shapes. 
These effects are further influenced by particle size, morphology, and interactions with substrates or the surrounding environment.

Non-uniformity is also affected by extrinsic contributions from organic residues, including aromatic polycyclic hydrocarbons introduced during processing and have been reported to act as additional luminescent centers in some hBN samples~\cite{neumann_organic_2023}.
These emitters contribute their own PL features and complicate the spectral response of the nanoparticles. 
Temporal instabilities such as blinking and bleaching, which are common in nanoscale fluorophores due to charge trapping or non-radiative recombination, further add to spectral variability.
These intrinsic and extrinsic effects account for the observed unpredictability in the PL spectra of hBN nanoparticles and highlight the challenges of achieving reproducible optical performance.

\begin{figure}[H]
    \centering
    \includegraphics[width=0.75\linewidth]{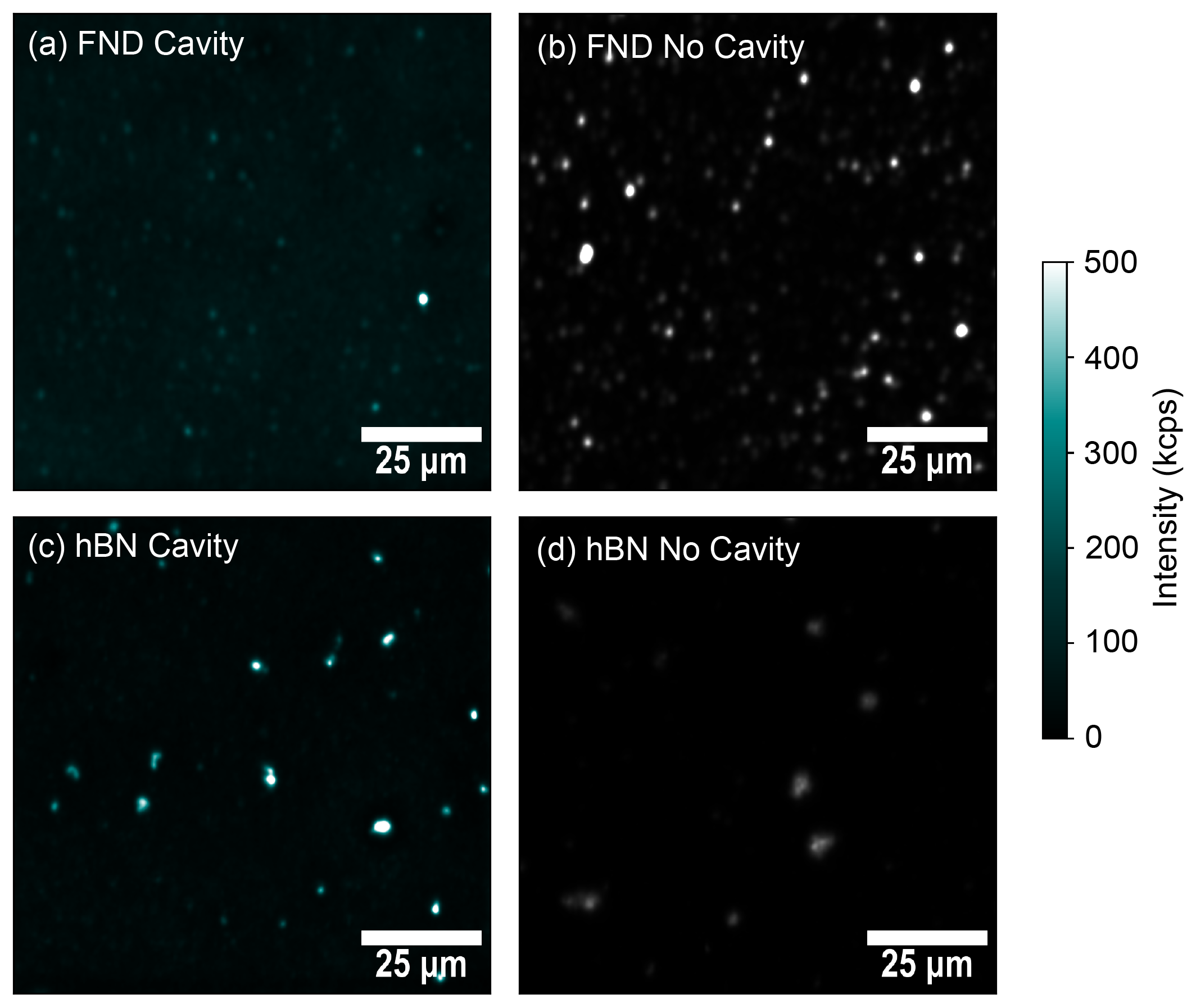}
    \caption{Confocal photoluminescence maps of quantum sensors in cavity and no cavity devices. (a) FND in the $\lambda_{res} = 650$~nm cavity device, (b) FND in no cavity device on the same substrate, (c) hBN in the $\lambda_{res} = 640$~nm cavity device, and (d) hBN in no cavity device on the same substrate.}
    \label{fig:PLmaps}
\end{figure}

\subsection{Exponential Decay Model}
\label{sec:fnd-lifetime}
To characterize the temporal decay dynamics of the emission, the time-resolved photoluminescence data were fitted using a bi-exponential model of the form

\begin{equation}
I(t) = A_{1} \exp\!\left(-\frac{t}{\tau_{1}}\right) + A_{2} \exp\!\left(-\frac{t}{\tau_{2}}\right),
\end{equation}

where $A_{1}$ and $A_{2}$ are the relative amplitudes, and $\tau_{1}$ and $\tau_{2}$ are the characteristic lifetimes of the fast and slow decay channels, respectively. This second-order exponential fit accounts for the presence of multiple recombination pathways or emissive states within the system, which cannot be adequately described by a single exponential. From the fitted parameters, an amplitude weighted average lifetime was calculated according to

\begin{equation}
\langle \tau \rangle = \frac{A_1\tau_1 + A_2\tau_2}{A_1 + A_2},
\end{equation}

which provides a representative metric for comparison across different spectra. This average lifetime captures the effective emission dynamics of the ensemble, while preserving sensitivity to both short- and long-lived components of the decay process.

\subsection{Quantum Microscope}
\begin{figure}[H]
    \centering
    \includegraphics[width=1\linewidth]{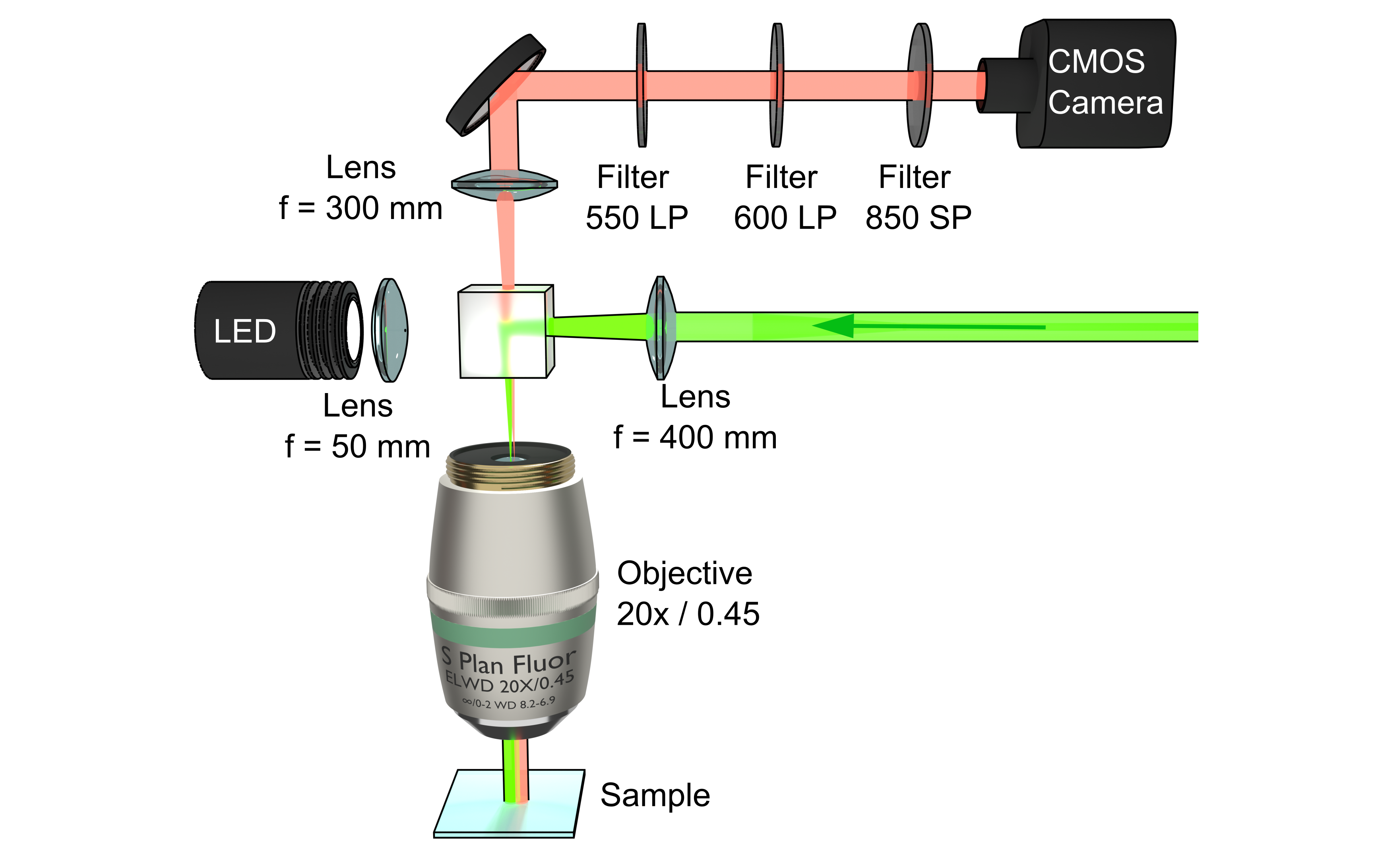}
    \caption{Schematic of the wide-field microscope used for fluorescence and bright-field imaging. A 532~nm excitation laser is focused into the back aperture of a 20×/0.45 NA objective. Emission is collected through the same objective, collimated with a 300 mm lens, spectrally filtered (500~nm LP, 600~nm LP, 850~nm SP), and imaged onto a CMOS camera. An LED coupled via a 50 mm lens and beamsplitter provides bright-field illumination for coarse alignment, while microwaves are delivered either via a wire loop or a PCB stripline depending on the substrate.}
    \label{fig:SI_widefield}
\end{figure}

The microscope utilized for ODMR cavity measurements is illustrated in Fig.~\ref{fig:SI_widefield}. Shown entering from the right hand side, excitation is provided by a 532~nm continuous-wave laser (Opus 2 W, Laser Quantum, Lastek, Australia), which passes through a 400 mm convergent lens and a 538~nm dichroic mirror (Semrock, USA). 
The beam was focused to the back aperture of the objective lens (S PLAN Fluor, 20x, NA = 0.45, Nikon, Japan).  
This results in a 50 \textmu m beam diameter at the sample with a power density of $\sim$0.5 mW \textmu m$^{-2}$ in the center. 
Fluorescence emission (red beam) was collected through the same objective and dichroic, then collimated using an f = 300 mm convergent lens. 
The emitted light was sequentially filtered using a 550~nm long-pass, a 600~nm long-pass, and an 850~nm short-pass filter (all from Thorlabs, USA) to block residual excitation and suppress background fluorescence outside the NV emission range. 
The image was collected with a CMOS camera (Zyla 5.5-W USB3, Oxford Instruments, UK). 
An LED emitting from 670~nm to 770~nm was also positioned before a 50 mm convergent lens and a right-angle beam splitter (R:T, 10\%:90\%).
The LED allowed bright-field imaging of the sample for coarse focusing.

Microwave (MW) driving was provided by a signal generator (Synth NV pro, Windfreak Technologies, United States) gated via an IQ modulator (TRF37T05EVM, Texas Instruments, United States) and amplified using a 50 W RF amplifier (HPA-50W-63+, Mini-Circuits, United States). 
A pulse pattern generator (PulseBlasterESR-PRO 500 MHz, SpinCore, United States) synchronized MW delivery, laser gating and camera acquisition. 
MWs were delivered to the samples via one of two methods dependent upon the sample.
For samples constructed on Si substrates, a wire loop antenna was positioned above the sample, delivering via the top mirror side.
Samples later built on quartz substrates were placed atop an exposed strip line waveguide in a printed circuit board (PCB), and microwaves were delivered to the FNDs through the quartz-mirror side.

\label{sec:Qmicroscope}

\subsection{ODMR Contrast of Cavities Fabricated on Silicon}
\begin{figure}[H]
    \centering
    \includegraphics[width=1\linewidth]{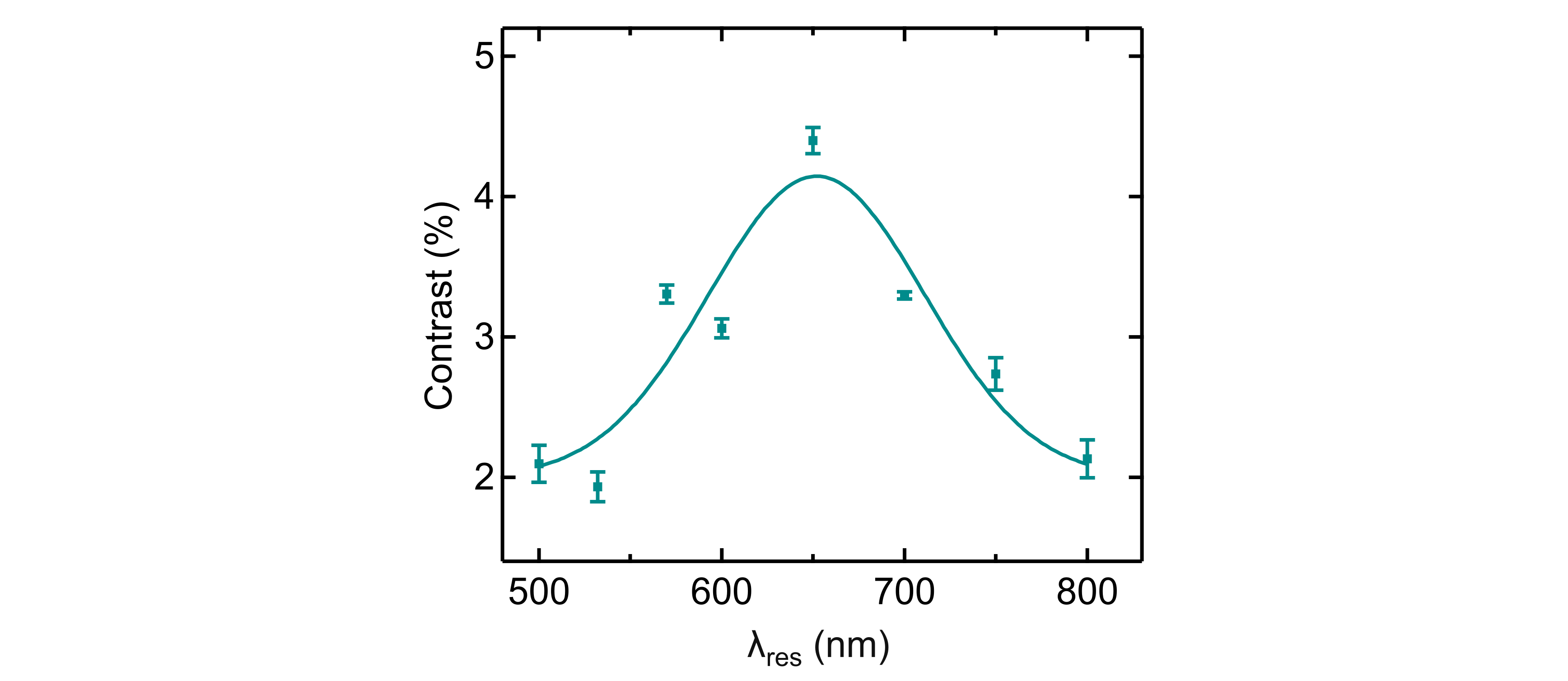}
    \caption{ODMR contrast of 100 nm FNDs in Si-substrate-cavities as a function of cavity resonance wavelength. Markers represent the ODMR contrast determined from averaged spectra across three regions of interest ($125 \times 125\ \mu$m) via a double Lorentzian fit. Error bars denote the fit uncertainty. The solid trace is a Gaussian fit to the data and a guide to the eye only.}
    \label{fig:SI_contrast_res}
\end{figure}
ODMR spectra for 100 nm FNDs in a cavities fabricated on Si substrates were acquired using the quantum microscope described in the preceding section.  
MW driving was supplied via a wire loop antenna to minimize absorption losses in the Si substrates. 
For each cavity, ODMR spectra were acquired for three regions of interest (125 x 125 \textmu m) and averaged. 
Figure \ref{fig:SI_contrast_res} plots the extracted ODMR contrast (markers) as a function of resonance wavelength. 
Each data point was obtained by fitting the ODMR spectrum to a Lorentzian function, with error bars representing the uncertainty in the fit.

\subsection{Normal Angle Reflectance Spectra of Cavity Devices on Quartz Substrates for Quantum Sensing}
\begin{figure}[H]
    \centering
    \includegraphics[width=0.5\linewidth]{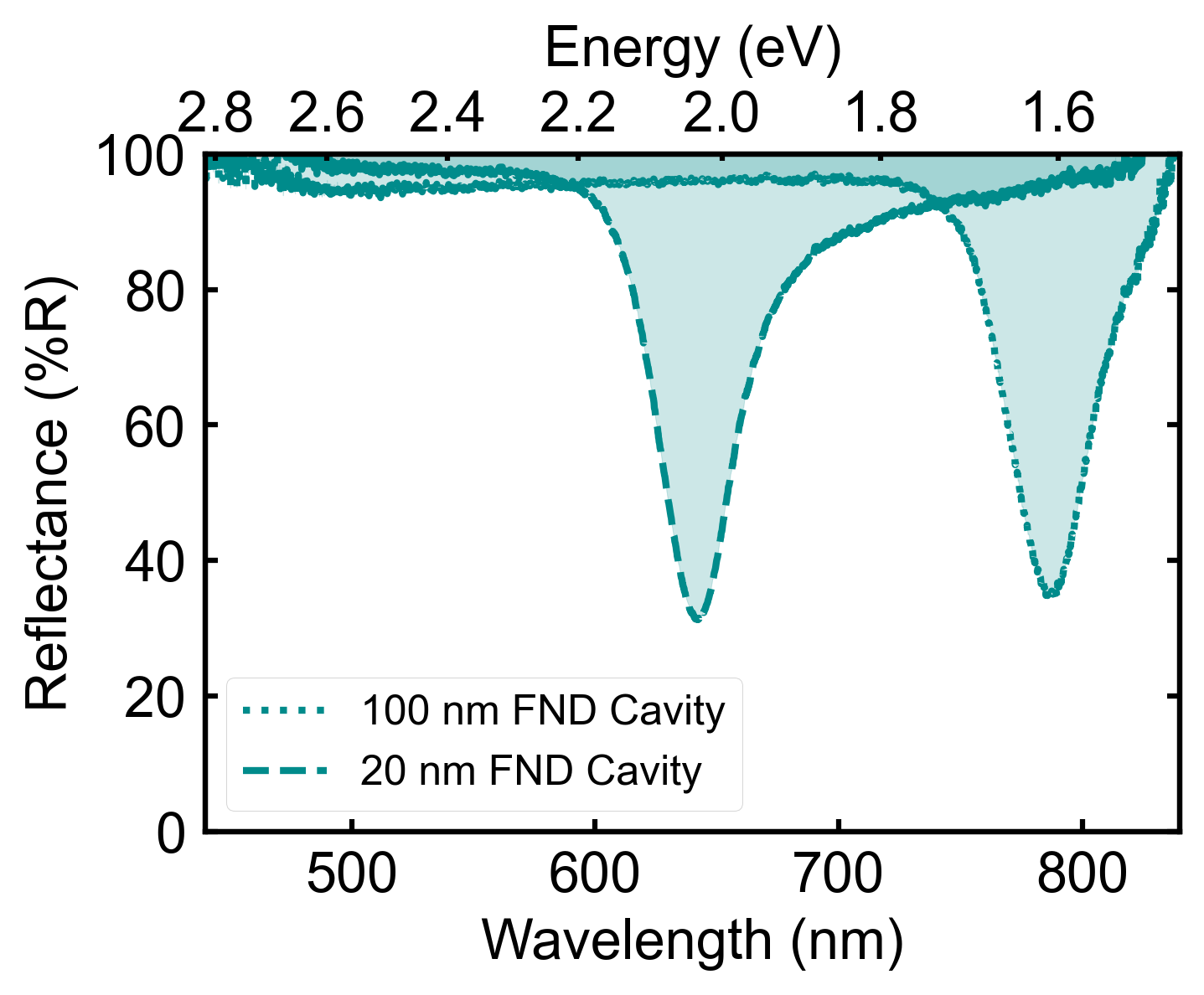}
    \caption{Normal angle reflectance spectra of the quantum sensing cavity devices.}
    \label{fig:fnd_q_reflectance}
\end{figure}

Normal angle reflectance spectra, used to determine the experimental $\lambda_{res}$ values for quantum sensing devices fabricated on quartz substrates, are displayed in Fig.~\ref{fig:fnd_q_reflectance}. 
Measured vs. predicted peak positions are displayed in Table~\ref{tab:q_devices}.

\begin{table}[H]
    \centering
    \renewcommand{\arraystretch}{1.2}
    \begin{tabular}{ccc}
        \hline
        Device Type & Predicted Resonance $\lambda_{0}$ (nm) & Actual Resonance $\lambda_{res}$ (nm)  \\
        \hline
        \hline
        PVP (100~nm FND-doped) & 650 & 788 \\
        PVP (20~nm FND-doped) & 650 & 642 \\
        \hline
        \hline
    \end{tabular}
    \caption{Cavity device or quartz parameters for quantum sensing. Predicted resonance is based on the transfer matrix model in Eq.~\ref{eq:tmm}, and actual resonance is determined according to the approximation in Eq.~\ref{eq:cav_approx}.
    }
    \label{tab:q_devices}
\end{table}

\subsection{FND Spectra from Cavities Fabricated on Quartz}
\begin{figure}[H]
    \centering
    \includegraphics[width=\linewidth]{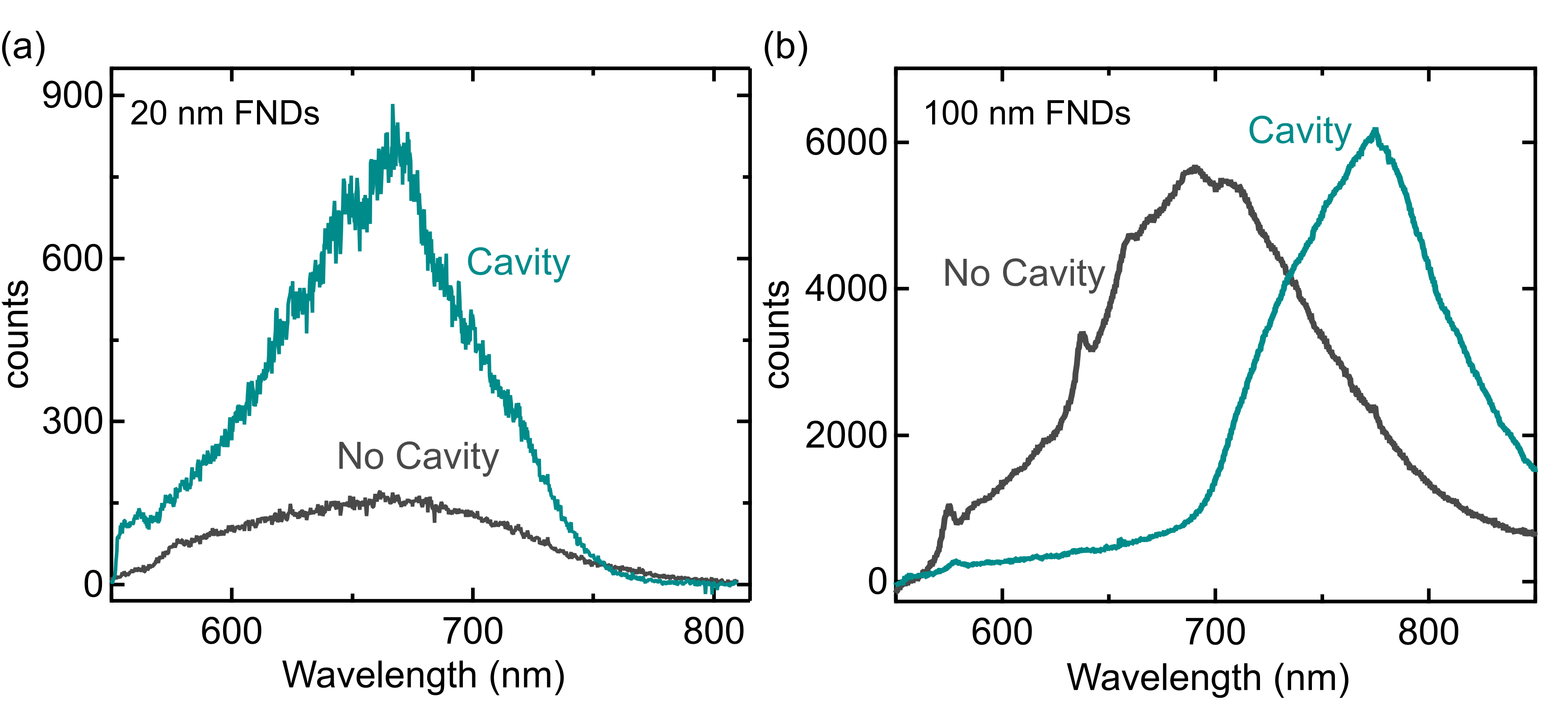}
    \caption{Spectra from particles inside (teal) and outside (grey) cavity regions for the 20~nm particles and 100~nm particles on quartz substrates. Both data sets were acquired using a confocal scanning microscope and scanning a 20 x 20 \textmu m randomly selected region. (a) Spectra from the 20~nm particles were acquired by conducting a long exposure of the spectrometer for the entire duration of the 20 x 20 \textmu m scans. (b) Spectra for the 100~nm particles were acquired by scanning a 20 x 20 \textmu m region and acquiring the spectra of 15 particles at random and averaging them together.}
    \label{fig:SI_spectra}
\end{figure}

Average PL spectra from FNDs inside and outside the cavity regions on quartz substrates are shown in Figure \ref{fig:SI_spectra}. 
The 20~nm FNDs (Fig.~\ref{fig:SI_spectra}a) exhibited a pronounced enhancement in PL brightness within the cavity regions, particularly in the NV$^-$ phonon sideband  region.
To obtain these spectra, PL spectra were acquired continuously during the acquisition of a 20 x 20 \textmu m confocal scan and averaged.
This method was utilized because the particle density of 20~nm FNDs was high enough that averaging over the scanned region yielded a representative spectrum without the need to isolate individual particles. 
In contrast, the 100~nm FNDs (Fig.~\ref{fig:SI_spectra}b) exhibited a narrowing of the emission spectra with an emission peak in the 700-800~nm region corresponding to the measured cavity resonance wavelength, as detailed in Table~\ref{tab:q_devices}, accompanied by a suppression of the PL at shorter wavelengths. 
\newline
Due to the lower density of 100~nm FNDs on the substrate, spectra were acquired by scanning a 20 x 20 \textmu m region and selecting 15 individual particles at random. 
These individual spectra were then averaged.

\subsection{FND Brightness from Cavities Fabricated on Quartz}
\begin{figure}[H]
    \centering
    \includegraphics[width=\linewidth]{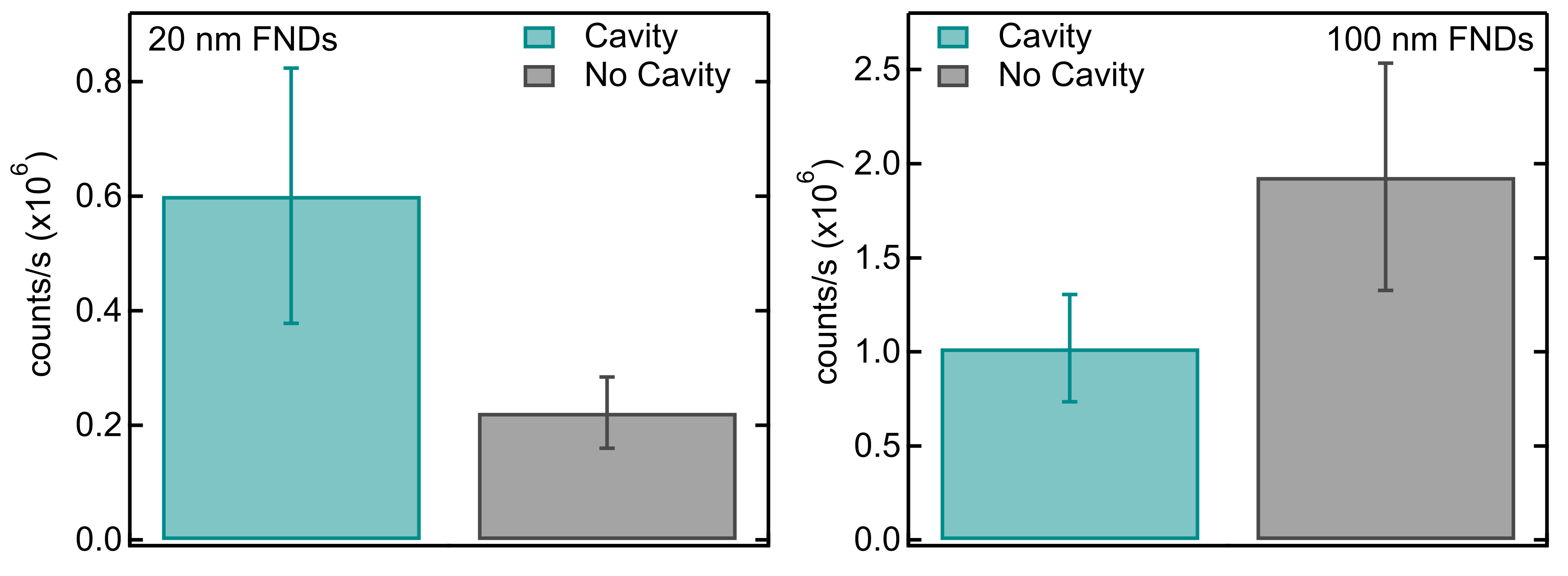}
    \caption{Brightness of particles inside (teal) and outside (grey) cavity regions for the 20~nm particles and 100~nm particles on quartz substrates. Data was acquired with a confocal scanning microscope by scanning a randomly selected region and observing the brightness of 15 particles each. The brightness of FND's were averaged together. The error bars show the standard deviation in the averaged FND brightness's}
    \label{fig:SI_brightness}
\end{figure}
The brightness of particles inside and outside of cavities was measured using a scanning confocal microscope with a 532~nm excitation laser at 300 \textmu W. 
On both samples a random region (20 x 20 \textmu m) was scanned on the cavity side and on the no cavity side, resulting in four scanned regions. 
From each scanned image, the brightness of 15 particles was measured and compared. 
On average 20~nm FNDs in the cavity were 2.7x brighter at $6\times10^5$ counts/s compared to those outside of the cavity. 
For the 100~nm FNDs the opposite trend was observed, being 0.4x dimmer inside the cavity than those outside. 
This result coincides with the observed spectral data in Fig.~\ref{fig:SI_spectra}, which shows that while the 20~nm FNDs had a large enhancement in brightness across the NV phonon sideband (PSB), the 100~nm enhancement resulted in a narrowing of the emission spectra in the PSB accompanied by a quenching of the PL at lower wavelengths. 
As a result, dimming from absorption and reflectance at the top cavity mirror may have a more significant effect in the 100~nm samples, where no clear broadband increase in PL intensity was observed. 
Thus, the apparent reduction in brightness of the 100~nm FNDs can be attributed to a combination of spectral redistribution and cavity-induced optical losses.

\subsection{Background Polymer Fluorescence in Polyvinyl Pyrrolidone}
Polyvinyl pyrrolidone (PVP) becomes slightly fluorescent in the visible after exposure to heat in the electron beam physical vapor deposition process. 
This emission readily photobleaches and is much dimmer than the emission from FNDs.
Polymethyl methacrylate (PMMA) shows no such fluorescence in the visible.
\begin{figure}[H]
    \centering
    \includegraphics[width=0.5\linewidth]{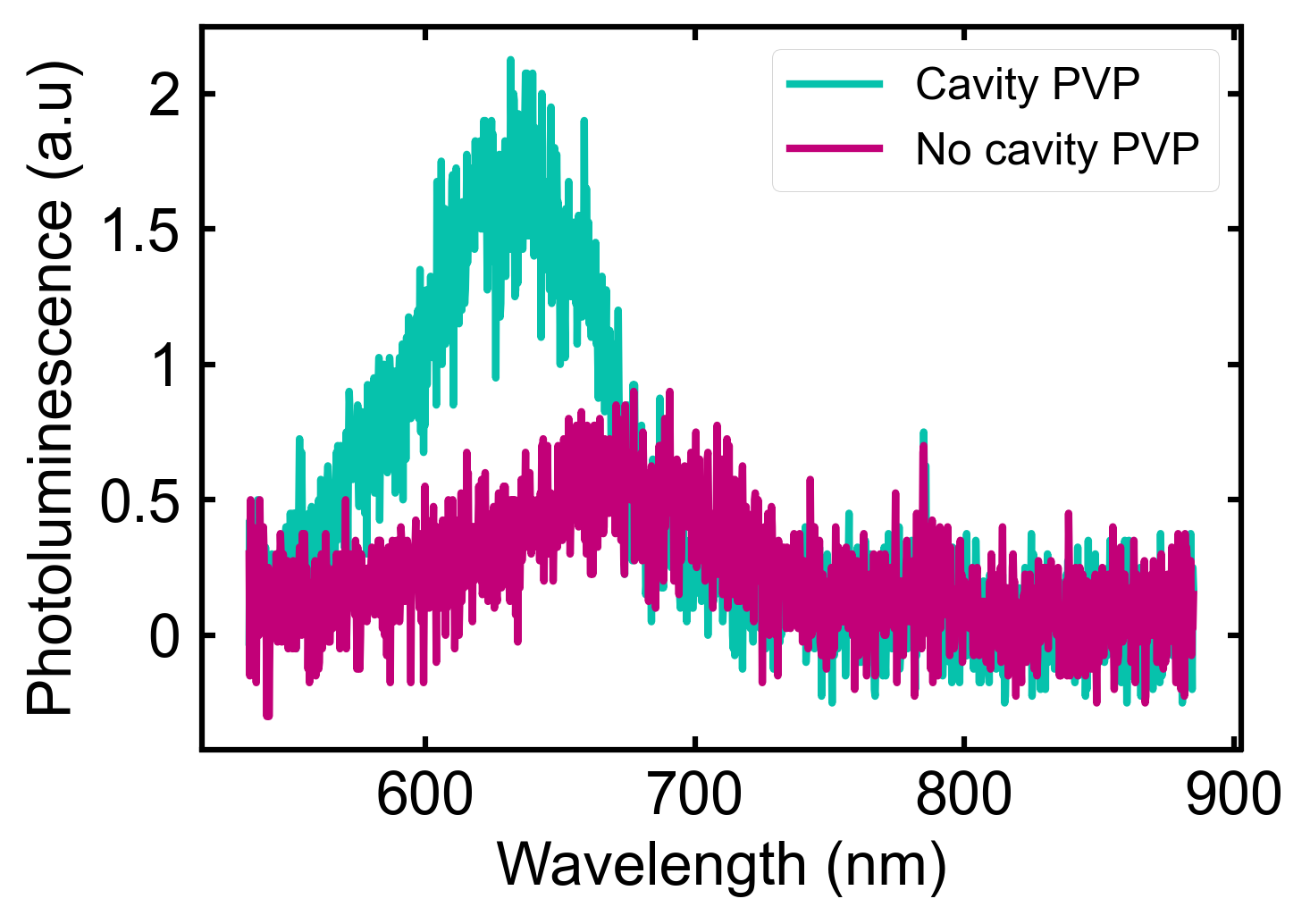}
    \caption{PVP polymer emission from the bare film inside and outside the cavity for a 650~nm resonant cavity.}
    \label{fig:polymer_pl}
\end{figure}


\bibliography{references_fnd.bib}

\end{document}